\magnification1200


\vskip 2cm
\centerline
{\bf  A brief review of E theory}
\vskip 1cm
\centerline{ Peter West}
\centerline{Department of Mathematics}
\centerline{King's College, London WC2R 2LS, UK}
\vskip 2cm
\leftline{\sl Abstract}
I begin with some  memories of Abdus Salam who  was my  PhD supervisor. After  reviewing  the theory of non-linear realisations and  Kac-Moody algebras, I explain how to construct the non-linear realisation based on the Kac-Moody algebra $E_{11}$ and its vector representation. I explain how this field theory  leads to  dynamical equations which contain an infinite number of fields defined on a spacetime with an infinite number of coordinates.  I then show that these unique dynamical equations, when truncated to low level fields and the usual coordinates of spacetime,  lead to  precisely the equations of motion of  eleven dimensional supergravity theory.   By taking different group decompositions of $E_{11}$ we find all the maximal supergravity theories, including the gauged maximal supergravities, and as a result the non-linear realisation should be thought of as a unified theory that is the low energy effective action for type II strings and branes. These results essentially confirm the $E_{11}$  conjecture given many years ago. 

\vskip2cm
\noindent

\vskip .5cm

\vfill
\eject

{\bf 0. Memories of Abdus Salam by one of his PhD students }
\medskip
I had the very good fortune to be a PhD student of Abdus Salam,  or 
Professor Salam as us students referred to him. I  began  in 1973 at Imperial College when supersymmetry was just beginning to be studied after the paper of Wess and Zumino [1]. Abdus  Salam was among a  small group of  largely Europeans who thought that supersymmetry  was interesting and had begun working on it in earnest. 

For my first year I did not see much of Professor Salam as I was taking my preparatory courses, but once the summer came I had finished my courses and so I went to see Professor Salam to find out  what I would work on. To my surprise he asked me what I wanted to do. I said that the infinities in quantum field theory were  very ugly and so I would like to work on general relativity which was more aesthetically pleasing. Rather than explain the flaws in  this naive approach, he suggested that there was not be so much to do in general relativity and that I might like to look at his very recent paper with John Strathdee in which they had discovered  superspace and also   super Feynman rules [2].   Within days I was captured by the ideas in this paper and began working on infinities in supersymmetric  theories. From the perspective of today  I realise  that I had been subject to his great charm and diplomacy,  a skill which he had used to such great effect all over the world. 

Once I had began research I never knew when I would see Professor Salam as he spent most of his time away from Imperial.  However, he would come about once a month. The first I knew that he was in the department was when, as I came in to work,  I would notice  that  his door  was slightly ajar. I would knock and he would welcome me  in.  He was always very cheerful,  friendly and seemed to have  time to talk as if he had all the time in the world. Little did I, as a student,  know of the many endeavours for the good of science in the third world that he was undertaking. 

At that time the problem was to  spontaneously break supersymmetry to find a realistic model of nature that was supersymmetric. Particles in a supersymmetric theory had the same mass and although one could spontaneously break  supersymmetry at the classical level [3] [4],  the pattern of masses it lead to was not consistent with nature. Professor Salam thought that  radiative corrections would break supersymmetry and lead to more promising results. Professor Salam, with John Strathdee,  who was in Trieste,  produced a series of models and it was my job to compute their one loop effective potentials and see if supersymmetry was spontaneously broken and what pattern of masses they lead to.  The first models  did not   work and as time went  on the models became more and  more   complicated involving  very many fields. If I had not completely finished  computing   with a given model by the time I next meet with Professor Salam it was not a problem,   there was always a much better model to look at instead. 

In such early days of supersymmetry there were no papers one could  look at to get up to speed with the technical difficulties, such a Fierz reshuffles,  that were required to  work  on supersymmetry. Fortunately Professor Salam's long term collaborator Bob Delbourgo had an office nearby and he provided me with all the technical help I needed. We also worked on some of the later models together, swopping rows and columns in matrices of large dimension in order to diagonalise then  so as to find the masses, which then turned out to be unsatisfactory. 

Eventually I realised that if supersymmetry was preserved at the classical level then the effective potential vanished in the most general $N=1$ theory invariant under rigid supersymmetry theory [5]. This meant that one could not  spontaneously break supersymmetry using perturbative quantum corrections,  although one could still hope that it was spontaneously broken by non-perturbative corrections. The problem of breaking supersymmetry in a natural way is  still largely unsolved. The result had another more favourable consequence,  as was pointed out by others [6], namely that  supersymmetry did solve the hierarchy problem, at least  technically. In a supersymmetric version of the standard model the Higgs mass would not be swept up to some large unified scale by quantum corrections as long as supersymmetry was not broken much above the weak scale. This in turn lead to the hope that supersymmetry  might be found at the LHC.  

Talking to Professor Salam you could not escape his great enthusiasm for physics; you came to understand that it was a lot of fun  to do physics and that it was good to work  in a very  relaxed and free thinking way. As became even clearer when  I later visited him in Trieste,  after I had my PhD,  Professor Salam could think of a vast number of ways   to proceed in the quest to find new things. He was always most interested in very new ideas and while not all of his ideas worked they  included many of the deepest ideas that have come to dominate the subject. As one of his students I was,  perhaps, able to absorb some of these qualities. Certainly,  it was due to him that I began working  on supersymmetry rather than on some uninteresting direction.

I end with an account of three  meetings with Abdus Salam that display his  warmth and   humanity. 

At  the end of my visit to Trieste the time came for me to leave for  the airport.  Salam realised that I would be travelling at the same time as the Italian Minister for Science, who was visiting the centre, and so he suggested we could share the same car to the airport. This was met with a frown by the organiser of the visit  who, no doubt correctly,  thought that a scruffy post-doc with a rucksack might dent the carefully created image that the centre wanted to portray. Of course I went by myself to the airport. 

I met Abdus Salam in his office in London a few days after he had won the Nobel prize. I asked him what was it like to win such a prize,  he reassured me that he was just the same. He then suggested that we go  for coffee in the common room in the old physics building at Imperial. To get there we had to go through a number of doors and he insisted that I go first through each door despite my protests. 

During the time that Salam was very ill there was a conference in his honour at Trieste, but he was not well enough to go to all the talks. I saw him sitting at the very back of the big auditorium. I asked if it would be alright to say hello, but I was told that he might not recognise me. Since this might be the last time I would see him I went anyway. I said hello,  he put up his hand and I shook it. He then immediately said how was Sue. Sue is my wife's name who he had meet only once many years before. 

\medskip
{\bf References for this section }
\medskip
\item{[1]} J. Wess and B. Zumino, {\it Supergauge transformations in
four dimensions}, Nucl. Phys. {\bf B70} (1974) 139; {\it A Lagrangian Model Invariant Under Supergauge Transformations}, Phys. Lett.
{\bf 49B} (1974) 52.
\item{[2]} A. Salam and J. Strathdee, {\it  On Superfields and Fermi-Bose Symmetry}, Phys. Rev. {\bf D11} (1975) 1521. 
\item{[3]} L. O'Raifeartaigh, {\it  	
Spontaneous Symmetry Breaking for Chiral Scalar Superfields},  Nucl. Phys. {\bf B96} (1975) 331. 
\item{[4]} P. Fayet, {\it  	Spontaneous Supersymmetry Breaking Without Gauge Invariance} Phys. Lett. {\bf 58B} (1975) 67.
\item{[5]} P. West, {\it Supersymmetric Effective Potential}; Nucl. Phys. {\bf B106} (1976) 219. 
\item{[6]} E. Witten, {\it Dynamical Breaking of Supersymmetry},  Nucl. Phys. {\bf B188} 513 (1981) 150. 

\vskip16cm

\eject
{\bf 1. Introduction}
\medskip
Quantum field theory, and  in particular quantum electrodynamics (QED),  was formulated  by  Heisenberg and Pauli  in 1929-30. However, they realised that there was a problem, all the  calculations in this new theory lead to infinities. Indeed, in  1930 Oppenheimer and Waller showed that the self energy of the electron was infinite.  The problem arose from  the sum over the undetermined momenta circulating in processes involving particles propagating in  loops. This lead to the feeling that quantum field theory was not a correct theory and that some deeper structures were required. However, just after the end of World War II many  distinguished physicists gathered for a meeting  at Shelter Island in  1947. Here  Kramers pointed out that the final result of the calculation,  say for the mass of the electron, was experimentally observed to be finite but the parameters of the theory were not measured and so one could try to  absorb the infinities in the parameters. Bethe, on the long train ride back to New York from Shelter Island, used this idea to calculate a quantum field theory (QED) correction to a certain  spectral line of hydrogen, the Lamb shift, which had just been experimentally measured. He found the correct result and the way was then clear to calculate more quantities  in QED, such as the magnetic moment of the electron,  which turned out to agree with subsequent experimental measurements with remarkable precision.  Confidence in QED was further boosted  when it was shown that the infinities could not only be absorbed for processes involving  simple Feynman diagrams but for all calculations in QED and the same  for certain other quantum field theories.  The 1950-51 papers of Salam were crucial to this result as they solved the problem of overlapping divergences. However, around this time Dyson showed that one could absorb the infinites for only a very limited class of quantum field theories. 
\par
However, there were also a significant  number of theorists who  believed that quantum field theory was not the correct framework to formulate the theory of the weak and strong nuclear forces.  In 1937   Wheeler and, independently,  in  1943 Heisenberg proposed  that one should work with measurable quantities rather than the many non-measurable quantities that appear in quantum field theory and in particular one should study the  S-matrix. It was this development that lead to string theory. 
\par
It is instructive to recall some of the very early developments of particle physics. The first  particles to be discovered  were  the electron,  the proton and then neutron (discovered in 1932), then  first glimpsed in cosmic rays were the  positron (1932), the muon (1936),  the  pions (1947) and  the  K mesons (1947). Subsequently  the neutrino (predicted by Pauli) was  found in 1956 in a nuclear reactor. With the advent of particle accelerators  many new particles were found. On the theoretical side, in 1932 Heisenberg  suggested that the neutron and proton might form an SU(2) symmetry multiplet if one neglected electromagnetic interactions.  Kemmer then  used this symmetry to write down an action involving fields for the proton, neutron and the three pions which were a triplet. 
\par
The problem for those that wanted  to formulate  the nuclear weak and strong forces using a  quantum field theory was that they had no principle to help them  determine the interactions of the many particles which were being discovered. In 1932 Fermi proposed an experimentally sucessful theory of weak interactions consisting of a four fermion interaction,  but it had infinities which could not be absorbed in the parameters of the theory and so it was not consistent with quantum mechanics.
In 1935 Yukawa  had proposed that there should exist some massive particles that would mediate the strong nuclear force and when the pions were discovered it was  initially thought that they were such particles.  However, if one viewed this exchange in the context of  a quantum field theory it required  a  large coupling constant and so it could not be analysed using perturbation theory.  Yang-Mills theory  was formulated in 1954 and,  independently by Shaw, a PhD student of Salam. This  did possess a principle that determined interactions and it had the above mentioned SU(2) symmetry in mind,  but it was difficult to see at that time how it could be compatible with the observed particles. While the idea that the nuclear weak and strong forces could be mediated by spin one bosons using Yang-Mills theory was being studied  there was no consensus on how to do this and what symmetry to use.
\par
A different approach to describe the behaviour of the almost massless pions  was put forward. It had been shown, by Goldstone, Salam and Weinberg,  that if a quantum field theory possessed a rigid symmetry with group $G$, that was spontaneously broken to a group $H$,  then one found the dimension of $G$ minus the dimension of $H$ massless particles (Goldstone theorem). 
Although it was not initially phrased in this way,  it was understood that the low energy dynamics of these massless particles was determined  by a non-linear realisation of  group $G$ with local subgroup $H$.    It was shown that if one took $G=SU(2)\otimes SU(2) $ and $H=SU(2)_{diag}$ then the dynamics predicted by the non-linear realisation agreed with the dynamics of the pions as it was measured in the particle accelerators provided one allowed for their small mass.   
\par
The great advantage of using  non-linear realisations was that it allowed the pioneers to find some of the symmetries of the theory without having to solve the much more difficult problem of what was the underlying theory, or indeed, even what new conceptual ideas it incorporated. The result was a  new appreciation of role of  symmetry as a principle to determine interactions and also  the importance  of spontaneous symmetry breaking.
These ideas  played a key role in  the development of the standard model by Glashow, Salam and Weinberg which is based on the Yang-Mills symmetry $SU(2)\otimes U(1)$ which is spontaneously broken.  
\par
Goldstone theorem also played an important  role in the development of the standard model. The idea to use  spontaneously broken symmetries to give mass to the spin one bosons  was thought not to be possible as Goldstone theorem would predict the presence of massless particles that were not observed. The crucial observation  was that if a symmetry was a local symmetry then Goldstone's theorem did not apply and one did not find the massless particles predicted by Goldstone's theorem.  Thus one could use spontaneous symmetry breaking to mediate the weak nuclear force as long as the symmetry was a local symmetry. Despite this it was not clear if the standard model was consistent as  it was known that the use of massive spin one  bosons to mediated  forces generally lead to infinities of a type that could not be absorbed in the parameters of the theory. However, it was found that if their masses arose  from the spontaneous breaking of a local Yang-Mills symmetry then  the infinities can be tamed and so the theory was consistent with quantum mechanics. 
\par
Perhaps the moral to be drawn from these developments is that the quest to understand nature at its deepest level requires more and more symmetry 
that is spontaneously broken. Also  demanding  consistency and mathematical beauty provides  a very powerful guide to finding the correct theory. The standard model emerged from combining special relativity and quantum theory 
in a way that lead to consistent theory, and in particular,  one whose predictions were free from infinities. It is also clear  that  progress on such a difficult problem  had to proceeded step by step and there was no hope that anyone could  guess the final theory, or even  the main underlying concepts in one single leap. 
\par
The history of particle physics I  have given has been partly derived from the books and papers in reference [0]. However, not having studied most of the original papers I may not have the emphasis correct in some places.   
\par
The quantisation of Einstein's theory also lead to the infinities mentioned at the beginning of this introduction, but unlike for QED and the standard model,  these can not be absorbed in the parameters of the theory. This can be seen to be a consequence of the fact that the vertices of this theory, unlike that of Yang-Mills theory contain momentum squared factors and this is turn can be traced to the  requirement that the theory be invariant under general coordinate transformations. These transformations contain a parameter which always comes with an associated derivative   unlike the transformations  of the Yang-Mills gauge field.  Thus it appears that the application of quantum theory and Einstein's theory of  general relativity in the context of point particle quantum field theory does not lead to a consistent theory of quantum gravity. 
\par
As we have mentioned, the disillusionment with point particle quantum field theory lead to S-matrix theory which in turn lead to string theory. The all loop scattering amplitudes  of the bosonic string were computed at a very early stage of string theory, although the integrand over the moduli of the Riemann surface was not evaluated in general [1,2]. These calculations lacked the contribution due to the ghosts as their necessity was not understood at this time, but this was provided later and  resulted in only in slight changes to the amplitudes.  The results can be written in a very compact way and one can  think that the simplest things about string theory are the amplitudes themselves. String theory does provide a  consistent theory of quantum gravity when viewed from a perturbative perspective in that the amplitudes are essentially free from the infinities referred to above.  The price is that one has an infinite number of particles corresponding to the vibrational modes that exist on the string length. However, there is no truly systematic way to compute non-perturbative effects in string theory and from  this viewpoint string theory is not complete. 
\par
One finds that closed strings  contain a graviton  and that the open strings contain gauge particles. Indeed,  at low energy,  open strings describe contain Yang-Mills gauge theory [3] while closed strings contain Einstein's theory [4]. Thus string theory has the potential to contain the particles responsible for the forces that we know. 
\par
The type II superstrings  are, by definition,  those that possess a  spacetime supersymmetry with a parameter that has thirty two components. There are two such theories called IIA and IIB. The massless particles  of the superstrings are essentially determined by  this supersymmetry and they are of necessity the states of the type II supergravity theories in ten dimensions which by definition have the same number of supersymmetries.    The low energy actions of the superstrings are by definition theories whose degrees of freedom are these massless particles. These theories are complete in that  they contain all effects at low energy,  including all the effects that are due to the heavier particles in intermediate processes. One can think of them as arising once the  integral over the heavier particles has  been carried out in the Feynman path integral of the underlying theory. 
\par
Given the very powerful character of spacetime supersymmetry in determining invariant  actions, or equations of motion,  it is natural to believe that   the low energy effective action of the type IIA  superstring is the type IIA supergravity [5] and the type IIB  superstring is the type IIB supergravity [6,7,8],  indeed this was the motivation for the construction of these supergravity theories. One might expect that  these supergravity theories should contain all the perturbative and non-perturbative superstring effects at low energy. As such they  have provided one source of knowledge about string theory that is complete  and  many  developments have  arisen from thinking about superstring theory from this perspective. In particular these supergravity theories contain solutions  that correspond to strings but also branes.   An obvious anomaly  was the existence of the eleven dimensional supergravity theory [9]. This theory does not possess a string solution but rather a two brane solution.  The IIA and IIB supergravity theories also contained brane solutions and one is lead to the expectation that strings and branes should be treated on a more  equal footing. 
\par
One of the most unexpected developments in supersymmetric theories was  that supergravity theories   possesses unexpected symmetries. Indeed the maximal supergravity theory in four dimensions was found to possess an $E_7$ symmetry [10].  More generally  the  maximal supergravity in $D$ dimensions were found to have an $E_{11-D}$ symmetry [11]. These studies did not include the IIB supergravity theory that was found to have a SL(2,R) symmetry [6]. These symmetries are associated with the scalar fields that these theories possess and indeed the dynamics of the scalar fields in these theories are described by a non-linear realisation of these groups. Such symmetries are broken by the presence of solitons and the quantised charges they possess and  it  was first proposed in the context of the heterotic string that such symmetries, when suitably discretised,  might be symmetries of string theory [12]. This was generalised to the type II superstrings and in particular the conjecture that the IIB superstring should have a SL(2, Z) symmetry [13]. One very interesting consequence of these symmetries was that the transformed the string coupling in such a way that they took the weak (small) coupling regime to the string (large) coupling regime of these theories [12,13].  
\par
The type II supergravity theories are connected by a number of relations. The dimensional reduction of the eleven dimensional supergravity theory on a circle leads to the IIA supergravity theory in ten dimensions, indeed this was how this latter theory  was constructed. We note that dimensional reduction on a circle preserves the number of supersymmetries. Further dimensional reduction leads to the unique maximal supergravity theories on nine and less dimensions. By a maximal supergravity we mean a theory that is invariant under a supersymmetry that has thirty two component parameters. The dimensional reduction of the IIB supergravity theory leads to the same nine dimensional theory; this must be the case as  the nine dimensional maximal supergravity theory is  unique. Thus there is a mapping between the IIA and IIB supergravity theories on a circle. These relations inherit into corresponding superstring theories. The strong coupling limit of the IIA string theory can be thought of as defining an eleven dimensional theory whose low energy limit is the eleven dimensional supergravity theory [14]. The relations between the IIA and IIB superstring theories on circles is an example  T duality transformation. 
\par
These ideas have come to be known as M theory but, as is rather clear, this is not a theory rather it is a set of relations between the  different theories.
\par
The problem with including branes as elements in the underlying theory is that unlike the string there is very little known about how to quantise branes. While  there has been progress on formulating an action for multiple coincident M2 branes  and the use of open strings has allowed us to understand some properties of D branes  one does not even know the quantum states of a single brane and the problem of scattering amplitudes for branes is still very far from being solved. Thus progress in  string theory has lead us into some kind of no mans land in that  we realise that we understand very little about what the underlying theory could be. 
\par
In this review we will explain that all the maximal supergravity theories, that is, the low energy effects of the superstring theories, can  be unified in a single theory that contains a very large new symmetry,  the Kac-moody algebra $E_{11}$. This theory is a non-linear realisation of the semi-direct product of $E_{11}$ with its vector representation denoted $E_{11}\otimes_s l_1$.This field theory is similar to that used to formulate the dynamics of pions mentioned above, but it differs in that it automatically contains a spacetime as part of the group structure. This is a first step that one can hope may be used to  determine some of the properties of  the underlying theory of string and branes. 
\par
We begin by giving a review of the theory of non-linear realisations in section two. Section three contains a short account of Kac-Moody algebras followed by  section four  gives some of the historical motivation for the idea that the underlying theory of strings and branes has an $E_{11}$ symmetry. Section five constructs  the Kac-Moody algebra $E_{11}$ and its representation which of most interest to us, the $l_1$, or vector,  representation. Section six  contains the construction of the non-linear realisation of $E_{11}\otimes_s l_1$ non-linear realisation and the eleven dimensional dynamics it predicts. Section seven explains how the theories in $D$ dimensions emerge from this non-linear realisation and section eight shows that  the non-linear realisation of $E_{11}\otimes_s l_1$ is a unifying theory in that it contains the many different  type II maximally supersymmetric theories. Section nine is a discussion of the meaning of the results. This review is an expanded version of the lecture given in Singapore. 
\medskip
{\bf 2. Non-linear realisations}
\medskip
As mentioned in the introduction the theory of non-linear realisations  was once  well known  but  this knowledge  has largely been lost, and  worse still,  been replaced by misunderstandings. As a result in this section  we will review the theory of non-linear realisation. The data required to specify a  non-linear realisations is a group $G$ with a choice of  subgroup $H$. The non-linear realisation of a group $G$ with local subgroup $H$  is, by definition,  constructed out of a group element $g\in G$ which is subject to the transformations 
$$
g\to g_0 g, \ \ \ g_0\in G,\ \ {\rm as \  well \  as} \ \ \ g\to gh, \ \ \ h\in H
\eqno(2.1)$$
The group element $g_0 \in G$ is a rigid transformation, that is, it is  a constant,  while $h\in H $  is a local
transformation, that is,  like $g$ it depends on the  space-time that  the theory possess. The spacetime may be introduced by hand,  as was the case for the original use of the non-linear realisation used in particle physics,  or it may be introduced  as part of the construction by including  corresponding generators that belong to the group $G$. This latter case  is the one of interest to us in this paper. Clearly we can use the local transformation $h$ to gauge away part of the group element $g$. 
\par
We now take the group $G$ to have a particular form, that is,  it is the semi-direct product of a group $\hat G$ with one of its representations $l$; we denote this by $\hat G \otimes_s l$. We denote the generators of $\hat G $ by $R^\alpha$ and for each element of the $l$ representation we introduce a corresponding generators $l_A$. Then the algebra 
$\hat G \otimes_s l$ can be written as 
$$
[R^\alpha , R^\beta ]= f^{\alpha\beta}{}_{\gamma} R^{\gamma}
\eqno(2.2)$$
$$
[R^\alpha , l_A ] = - (D^\alpha )_A{}^B l_B
\eqno(2.3)$$
The first equation is just the Lie algebra for $\hat G$ and in the second equation the matrix $(D^\alpha )_A{}^B$ is the matrix representation of the $l$ representation. One can verify that it satisfies the Jacobi identity by virtue of this fact. The commutators of the $l$ generators are restricted by the Jacobi identity. The simplest consistent choice is to take them to commute but  we will leave them unspecified for the time being. 
\par
The reader is very familiar with the notion of the semi-direct product as the Poincare group $P$ in $D$ dimensions can be written as $P= SO(1,D-1) \otimes_s T^D$ where $T^D$ are the translations generators corresponding to the vector representation of $SO(1,D-1)$. If we denote the spacetime translations by $P_a$ and the Lorentz rotations by $J_{ab}$, with $a,b,\ldots =0,  1,2 , \ldots , D-1$ then the   Lorentz  algebra is given by 
$$
[J_{ab} , J_{cd} ]= \eta_{bc} J_{ad}- \eta_{ac} J_{bd}-\eta_{bd} J_{ac}+ \eta_{ad} J_{bc}
\eqno(2.4)$$
 while  equation (2.3)  is written, for this case,  as 
$$
[J_{a b} , P_c ]= -\eta _{ac} P_b + \eta _{bc} P_a 
\eqno(2.5)$$
\par
The group element $g$ of $\hat G \otimes_s l$ can be written in the form 
$$g= e^{x^A l_A} e^{A_\alpha R^\alpha} \equiv g_l g_A
\eqno(2.6)$$
where $x^A$ and $A^\alpha$ parameterise the group element and in the second equation $g_A$ and $g_l$ involves the generators of $\hat G$ and the $l$ representation respectively. We will interpret the $x^A$ as the coordinates of a spacetime and the  $A^\alpha$ as fields that live on this spacetime, that is, they depend on the coordinates $x^A$. 
\par
The dynamics of the non-linear realisations is just a set of equations that are invariant under the transformations of equation (2.1). To understand why the non-linear realisation leads to  equations of motion one just has to realise that the group element $g$  of equation (2.1) contains  the fields of the theory which depend on the generalised space-time. As a result when one finds a set of quantities, constructed out of the group element $g$,  that  is,  invariant under the transformations of equation (2.1) one is necessarily constructing an equation of motion for the fields of the theory. Hence the non-linear realisation leads to  dynamical equations for the fields which are either unique,  or almost  unique,  provided one specifies the number of derivatives involved. 
As with every application of any symmetry one has to specify the number of spacetime derivatives the action should contain.  Non-linear realisations  are  a bit different to the more familiar situation where one has some fields that transform linearly under a symmetry as in the case  of the non-linear realisation the symmetry and the fields are very closely linked and it is this that leads to the prediction of the dynamics in such a precise way. 
\par
We now consider three types of non-linear realisation, one that leads just to a spacetime, one that leads to fields that depend on a spacetime that is introduced by hand and finally one that leads to a spacetime and fields that depend on this spacetime. We denote these as types I, II and III. 
\medskip
{ \bf 1. Type I } 
\medskip
Let us first consider the case that the local subgroup $H= \hat G$ and in this case we can write the group element in the form 
$g= e^{x^A l_A} $
as the second factor in the group element involving the group $\hat G$ 
can be gauged away using the local $H$ transformation of equation (2.1). Thus in this case we are just left with the coordinates $x^A$ and there are no fields. If we take $G$ to be the Poincare group, that is, $\hat G= SO(1,D-1)$ and $H=  SO(1,D-1)$ the group element is $g= e^{x^a P_a}$ and the transformations resulting from the rigid transformation are the Poincare transformations of Minkowski spacetime. 
\par
Another classic example is to take $G$ to be the super Poincare group in four dimensions with one supercharge $Q_\alpha$  and $H=SO(1,D-1)$. The super Poincare group is a semi-direct product of the Lorentz group and its representation consisting of $P_a$ and $Q_\alpha$. We note that  in this case  the elements of the albeit reducible $l$ representation no longer commute with themselves.  The group element can be chosen to be of  the form $g= e^{x^a P_a} e^{\theta ^\alpha Q_\alpha}$ and the rigid transformations are those of superspace first found in the classic paper of Salam and Strathdee, reference [2] in the first section. 
\par
Type I non-linear realisation contain no fields and are just  the cosets $G/  H$ found in elementary mathematics books on group theory. 
\par
The Cartan forms for this type of non-linear realisation are given by 
$$
{\cal V} = g^{-1} d g = dx^\Pi e_\Pi {}^A l_A + dx^\Pi \omega _\Pi {}_{,\alpha} R^\alpha 
\eqno(2.7)$$
By studying the transformations of the objects  $e_\Pi {}^A$ and $\omega _\Pi {}_{, \alpha} $ under the non-linear realisation one finds that they can be taken to be the  vielbein and spin connection of the coset space G/H. It is this interpretation that encourages the use of the local indices $\Pi , ..$ rather than the tangent indices $A, \ldots $ according to whether they transform under local $H$ transformations or rigid $g_0$ transformations induced from such transformations on  the coordinates. The tangent space of the coset has tangent group $H$ and it is easy to find that $e_\Pi {}^A$ and $\omega _\Pi {}^\alpha $  transform under the local group $H$ as they should. 
\medskip
{\bf 2. Type II} 
\medskip
We now consider a second kind of non-linear realisation which involves taking no generators in the $l$ representation, that is $G=\hat G$ and $H$ is a subgroup of $ G$. The group element takes the form 
$g= e^{A_\alpha R^\alpha}$. So far we have no spacetime but, by hand,  we introduce a spacetime with coordinates $x_A$ simply by taking the fields $A_\alpha$ to depend on these coordinates. We note that the coordinates are dummy variables and, in this case, have no relation with the  generators $G$. Local in this case means that the group element $g$ and the local transformations $h$ of equation (2.1) depend on the coordinates $x_A$. One can use the local symmetry to choose the group element to be of a particular form and so set to zero some of the fields $A_{\alpha}$. Indeed the number of fields one can set to zero is the dimension of $H$ leaving the dimension of $G$ minus the dimension of $H$ fields. A fact that is consistent with Goldstone's theorem.  
\par
Our problem is to find the dynamics that is invariant under the transformations of equation (2.1). The usual method is to construct the Cartan forms 
$$
{\cal V}= g^{-1} d g=P+Q
\eqno(2.8)$$
where $Q$ belongs to the Lie algebra of $H$ and $P$ contains only the remaining generators in $G$. Considering the rigid transformations of equation (2.1) we see that the Cartan forms are invariant under these transformations. However, under the local transformations they transform as 
$$
{\cal V}^\prime= h^{-1} {\cal V} h +  h^{-1}dh
\eqno(2.9)$$
Clearly the $P$ part of the Cartan form transforms covariantly, that is, as $P^\prime =  h^{-1}P h$. If we demand that the action we seek has only two derivatives then it is of the form 
$$
\int d x_A Tr (P^2) 
\eqno(2.10)$$
where we have chosen the generators of $G$ to be in a particular representation, indeed any matrix representation will do. The number of possible terms in the action one can write is determined  by the way the adjoint representation of $G$ decomposes into the representations of $H$. If there is only one representation in addition to the adjoint representation of $H$ then the action is unique. The general theory for this  type of non-linear realisation was given in the  classic papers [15]. A more extensive review of this type of non-linear realisation can be found in section 13.2 of reference [16].  
\par
As we have mentioned above the maximal supergravity theories have some symmetries associated with the scalars. In fact the dynamics of the scalars in these supergravity theories is just the non-linear realisation of the corresponding symmetry group. In particular the IIB supergravity theory has two scalars and their dynamics is the non-linear realisation of SL(2,R) with local subgroup SO(2) [6], while the maximal supergravity theory in four dimensions has 70 scalars that belong to the non-linear realisation of $E_7$ with local subgroup SU(8) [10]. In general the scalars in the maximal supergravity theory in $D$ dimensions, for $D\le 9$, belong to the non-linear realisation $E_{11-D}$ with a local subgroup which is the maximal compact subgroup of $E_{11-D}$. 
\par
It was a  type II  non-linear realisation that was used to account for the pion dynamics, discussed in the introduction,  by taking  $G=SU(2)\otimes SU(2) $ and $H$ to be the diagonal SU(2) subgroup. The low energy dynamics is uniquely determined.  
\par
\medskip
{\bf 2. Type III} 
\medskip
Finally we give an account of the type of non-linear realisation used in this talk. Now we consider no restriction and so $G= \hat G \otimes_s l$ and the local subgroup $H$ is a subgroup of $\hat G$. The group element has the form of equation (2.6) and we find a spacetime with coordinates which are in one to one correspondence with generators in the $l$ representation.  The fields $A_\alpha$ depend on the coordinates $x_A$ and the group element transforms as in equation (2.1).   
\par
Since the generators  in the $l$ representation belong, by definition,  to a representation of $\hat G$ we can write the transformations of equation (2.1) under the rigid  $g_0$ belonging to $\hat G$ act as   
$$ 
g_l^\prime = g_0 g_l g_0^{-1}  , \quad g_A^\prime = g_0 g_A 
\eqno(2.11)$$
An exception is when the rigid transformations  $g_0  \in l$ and in this case they  just give a shift the coordinates. 
While the local $h\in H$ transformations act as 
$$ 
g_l^\prime =  g_l   , \quad g_A^\prime =  g_A h
\eqno(2.12)$$
As a result the local subalgebra transformations only change the fields and leave the coordinates alone. 
\par
To construct the dynamics we consider the Cartan forms which now take the form 
$$
{\cal V}\equiv g^{-1} d g= {\cal V}_A+{\cal V}_l, 
\eqno(2.13)$$
where 
$$
{\cal V}_A=g_A^{-1}dg_A\equiv dz^\Pi G_{\Pi, \underline \alpha} R^{\underline \alpha},
 \eqno(2.14)$$
belongs to the Lie algebra $\hat G$ and are the Cartan forms for $\hat G$, while  
 the part  that contains the  generators of the $l$ representation is given by 
 $$
{\cal V}_l= g_E^{-1}(g_l^{-1}dg_l) g_E= g_E^{-1} dz\cdot l g_E\equiv 
dz^\Pi E_\Pi{}^A l_A  
\eqno(2.15)$$
\par
While  both ${\cal V}_E$ and ${\cal V}_l$ are invariant under rigid transformations,   under local transformations of equation (2.5) they transform as the 
$$ 
{\cal V}_E\to h^{-1}{\cal V}_E h + h^{-1} d h\quad {\rm and }\quad 
{\cal V}_l\to h^{-1}{\cal V}_l h 
\eqno(2.16)$$
\par
Type III non-linear realisations were not as well studied as type II in the old days. However, Isham, Salam and Strathdee worked out in detail the non-linear realisation of the conformal group in four dimensions with the local subgroup being the Lorentz group [17]. Borisov and Ogivestsky considered the non-linear realisation of $GL(4)\otimes _sT^4$ [18]. In this case the dynamics was not unique but one could choose the undetermined coefficients so that  it lead  Einstein's gravity. A review of this calculation in $D$ dimensions can be found in section 16.2 of reference [16] which also develops the theory of type III non-linear realisation further as was done in   the $E_{11}$ papers referenced later on in this review. An  early review which also contains a discussion of these type III non-linear realisations can be found in reference [19]. 
\medskip
{\bf 3. Kac-Moody algebras}
\medskip
 In this section  we will explain how Kac-Moody algebras were  discovered [20, 21] and by doing so  give some insight into  what they are. We will gloss over many important points, however, the reader can read  a detailed and   pedagogical account of Kac-Moody algebras in chapter 16 of reference [16]. Group theory emerged from the study of the roots of polynomial equations, however, physicists are more used to thinking of groups as sets of matrices. Given a group it was found that one could reconstruct the   part connected to the identity by considering the Lie algebra. It also turns out that all finite dimensional Lie algebras can be constructed from a subset of Lie algebras that are finite dimensional and semi-simple and so we work just with these. The precise meaning of semi-simple can be found in section 16.1 of refence [16]. The Lie algebra, denoted $E$,  contains a set of commuting generators which we denote by $H_i, \  i=1,2,\ldots ,r$ where $r$ is by definition the rank of the algebra. This Abelian algebra is called the Cartan subalgebra. 
We can now diagonalise the remaining generators $E_{\alpha}$ with respect to 
the Cartan subalgebra, that is, we write the commutators of all the other generators in the Lie algebra $E$ with those of the  Cartan subalgebra generators in the form  $[H_i, E_{\alpha}]= \alpha_i E_{\alpha}$.  In doing this  we find a set of vectors $\alpha_i$ called the roots. A basis for the roots is called  the simple roots, and we denoted them as $\alpha_a, \  a=1,2,\ldots ,r$. Given the simple roots  we can construct their  scalar products to form the Cartan matrix which is defined by 
$$
A_{ab} = {2 (\alpha_a , \alpha_b) \over (\alpha_a , \alpha_a) }
\eqno(3.1)$$
\par
The classification of Lie algebras is usually carried out when the Lie algebra $E$ is considered to be over the complex numbers. However, it turns out that the Cartan matrix is real and has integer values. The Dynkin diagram consists of $r$ dots that are connected by a set of lines which are drawn according to the Cartan matrix using a set of  rules. The rules are such that given the Dynkin diagram one can deduce  the Cartan matrix uniquely. 
\par
Killing,  together with later work by Cartan, found that all the Lie algebras they knew   lead to Cartan matrices with the properties 
$$
A_{ab}\le 0\  {\rm  if}\  a\not= b
\eqno(3.2)$$
$$
{\rm if} \ A_{ab}=0 \ {\rm  then}\  A_{ba}=0
\eqno(3.3)$$
 $$
v^a A_{ab} v^b \ge 0 \ {\rm for \ any\  real \ vector}\  v^a
\eqno(3.4)$$ 
By construction $A_{aa}=2$ and the positive definite nature of the Cartan matrix implies that the off diagonal entries can only take the values $0,-1,-2, -3$ 
\par
Killing  looked at the possible list of Cartan matrices that satisfied the above properties and he found that there were some that did not corresponding to any Lie algebra that he knew. By finding the Lie algebras that lead to these new Cartan matrices he discovered some new algebras which were the exceptional algebras  $F_4, G_2, E_6, E_7$ and $E_8$. 
\par
In the above discussion we started from a Lie algebra and found a Cartan matrix.  However, in the 1950's Serre showed that one could go the other way around, that is, start from the Cartan matrix and reconstruct the corresponding Lie algebra. He introduced $3r$ generators $E_a, F_a$ and $H_a$.   The Lie algebra was just given by  all commutators of these generators subject to certain relations between these commutators that are completely specified by the Cartan matrix. We will not give them here but they can be found in section 16.1  of reference [16].  Once the Lie algebra has been constructed one can identify  the $H_a$ as  the Cartan subalgebra generators in a different basis, the $E_a$ as   the generators corresponding to the simple roots $\alpha_a$ and the $F_a$ as  the generators corresponding to the  roots $-\alpha_a$. 
\par
Kac-Moody algebras were discovered in 1969 [20, 21]. As Serre advocated we start from a Cartan matrix and construct the Lie algebra in the same way and subject to the same relations. However, we now allow Cartan matrices that obey only equations (3,2) and (3.3) but not necessarily  equation (3.4).  
\par
Clearly if equation (3.4) holds then the Kac-Moody algebra constructed is one in the Cartan-Killing list of Lie algebras, that is, the list of finite dimensional semi-simple Lie algebras. When the Cartan matrix is positive semi-definite with only one zero eigenvalue one finds that the construction leads to the already known and well understood affine Lie algebras. However in general one finds a vast new class of algebras whose properties are largely unknown. In particular one does not know a listing of the generators for even one of these new Kac-Moody algebras.  
\medskip
{\bf 4. Historical motivation for an $E_{11}$ symmetry}
\medskip
 As we have mentioned already, the $E_7$ symmetry of the maximal supergravity theory in four dimensions is associated with the seventy scalars whose dynamics is just a type II non-linear realisation, discussed in section two,   for the group $E_7$ with local subgroup $SU(8)$  [10].   As discussed in section two we may use the local symmetry to choose  part of the group element $g$ used in the non-linear realisation. In particular we may use the local subgroup  $SU(8)$ to remove part of the group element $g$ and it turns out that  one can choose the group element to belong to the Borel subalgebra of $E_7$. This is consistent with the count $133-63 =70$.  As a  result   every scalar in the supergravity   theory arises in the non-linear realisation from  a generator in the Borel  subgroup of $E_7$.  Indeed this is the general pattern for  the maximal supergravity theories,  one can use the local transformations in $D$ dimensions to choose  the group element associated with the non-linear realisation, to which the scalars belong,  to be in the Borel subalgebra of the symmetry group $E_{11-D}$. This  is related to the fact that the local subgroups used in the non-linear realisations are the maximally compact subgroups of $E_{11-D}$,  or more technically the Cartan Involution invariant subgroup. 
\par
The $E_{11-D}$ symmetries that arise in $D$ dimensional maximal supergravity theories were universally thought to be a quirk of the dimensional reduction procedures used to obtain these theories. However,  it was shown  that the eleven  dimensional supergravity theory was a nonlinear realisation [22]. This theory has no scalars but by introducing the generators 
$K^a{}_b$,  $R^{a_1a_2a_3}$ and $R^{a_1\ldots a_6}$ corresponding to the graviton $h_a{}^b$, three form $A_{a_1a_2a_3}$  and six form $A_{a_1\ldots a_6}$ fields respectively,  and taking them to obey a suitable algebra ${\cal A}^{11}$,  one could construct a non-linear realisation that lead to the eleven dimensional supergravity theory. We note that these  generators  carry indices that transform under the spacetime transformations, unlike for the   non-linear realisations that occurs for the scalar fields.  Not every theory can be formulated as a  non-linear realisation and so  this result told  us something about eleven dimensional supergravity. However, the dynamics of this non-linear realisation was not unique and it contained some constants that had to be fixed by hand to the required values. 
\par
One motivation for this construction was the previously mentioned, and rather old,   result of Borisov and Ogievetsky [18] which showed that gravity in four dimensions could be formulated as  a (type III) non-linear realisation of $GL(4)\otimes_s l$ with local subgroup SO(1,3) where $l$ is the vector representation of $SL(4)$. As we have just mentioned the gravity sector in the eleven dimensional supergravity theory arose from the $K^a{}_b$ generators which belong to the algebra GL(11). 
\par
The algebra ${\cal A}^{11}$  that emerged from formulating the eleven dimensional supergravity theory as a  non-linear realisation   was not a finite dimensional algebra in the list of Cartan and nor was it a Kac-Moody algebra. This was to be expected as using this construction one only finds generators associated with the fields of the theory and this does not include generators for any local subalgebra. Indeed one should expect to find only the Borel subalgebra of some large algebra. However, if one demanded that this algebra ${\cal A}^{11}$  was contained in a Kac-Moody algebra then the smallest such algebra  was $E_{11}$ [23]. Motivated by this realisation  it was conjectured that the $E_{11-D}$ symmetries were not a quirk of dimensional reduction but that the exceptional symmetries found in the lower dimensional maximal supergravity theories were part of a vast $E_{11}$ symmetry of the eleven dimensional theory [23].  The price to pay for  
changing the algebra used in the  non-linear realisation to $E_{11}$ was that it lead to a theory that contains an infinite number of fields,  only the first few of which were those of eleven dimensional supergravity. 
\par
It is instructive to examine  how the above construction would have proceed for the four dimensional maximal supergravity.   This theory can also be formulated as a non-linear realisation. To do this one  introduces for each field of the theory a generator and adopts a suitable Lie algebra that  they satisfy which  can largely be found by requiring that the corresponding non-linear realisation gives the equations of motion of four dimensional maximal supergravity  theory. In particular for the scalar fields one introduces   generators, which carry  no Lorentz indices,  in the Borel subalgebra of $E_7$, however, we must also introduce  the  generators $K^a{}_b$ for the graviton and $R^{aN}$ for the vectors.  We can think of this as extending the non-linear realisation of the scalars to include the other fields and in so doing so we must introduce generators that transform non-trivially under spacetime transformations. We note that we only have some of the generators of the full algebra in particular we have only the Borel subalgebra of $E_7$ rather than the full $E_7$ algebra. Demanding that the  algebra be extended to a Kac-Moody algebra leads to the  $E_7$ algebra in the  scalars sector, but the algebra $E_{11}$ for the full theory. 
\par
In the above we have sidestepped  the question of how we are to introduce spacetime into the theory. Thinking about the gravity sector of the non-linear realisation it is apparent that we should consider a type III non-linear realisation, that is, include generators in the algebra which lead to the coordinates of spacetime,  rather than the type II non-linear realisation used for the scalars. In the first papers on $E_{11}$ one just introduced the space time translation generators $P_a$ even though it was clear that this could only be part of the solution as it was not an $E_{11}$ 
covariant introduction. The correct way to introduce spacetime is to introduce generators corresponding to a representation of $E_{11}$, the $l_1$ representation,  which generalises the spacetime translations to include an $E_{11}$ multiplet of generators,  and take the algebra used in the non-linear realisation to be the semi-direct product of $E_{11}$ and this $l_1$ representation [24]. 
\par
The above method of proceeding has an analogy with the original use of non-linear realisations in particle physics. The analogue of pion dynamics is the maximal supergravity theories as these are thought to contain the low energy dynamics of strings and branes. The algebra  $SU(2)\otimes SU(2)$ is replaced with $E_{11}$. Of course the theory of pion dynamics was finally understood after  the introduction of quarks and the SU(3) gluon gauge theory.  However, as we explained in the introduction this approach to pion dynamics played a key role in unravelling the correct theory, that is, the standard model.   


\medskip 
{\bf 5 The $E_{11}$ algebra,  its vector representation and the $E_{11}\otimes_s l_1$ algebra}
\medskip
The Dynkin diagram of the Kac-Moody algebra $E_{11}$  is given by 
$$
\matrix{
& & & & & & & & & & & & & & \otimes & 11 & & & \cr 
& & & & & & & & & & & & & & | & & & & \cr
\bullet & - & \bullet & - & \bullet & - & \bullet & - & \bullet & - & \bullet & - & \bullet & - & \bullet & - & \bullet & - & \bullet \cr
1 & & 2 & & 3 & & 4 & & 5 & & 6 & & 7 & & 8 & & 9 & & 10 \cr
}
$$
As with any Kac-Moody algebra we do not know all the generators of $E_{11}$.  However, one can gain some understanding of the $E_{11}$ algebra by considering  the   decomposition of $E_{11}$ into representations of SL(11). In the Dynkin diagram this corresponding to deleting node eleven so as to  leave the algebra SL(11).This is the meaning of the cross in the $E_{11}$ Dynkin diagram.   The  decomposition leads to a set of generators which belong to representations of SL(11) and can be classified according to a level. The level  is the number of the number of up minus down SL(11) indices divided by three and it is preserved by the commutators of $E_{11}$. We will denote the generators of $E_{11}$ by $R^{\underline \alpha} $ and those of the $l_1$ representation by $l_A$. 
The positive level generators are given by [23]
$$
K^a{}_b, \ R^{a_1a_2a_3}, \ R^{a_1a_2\dots a_6} \ {\rm and }\  R^{a_1a_2\ldots a_8,b}, \ldots 
\eqno(5.1)$$
where the generators $R^{a_1a_2a_3}$ and  $R^{a_1a_2\dots a_6} $ are totally antisymmetric in their indices, while the next  generator obey the constraints 
$R^{a_1a_2\ldots a_8,b}= R^{[a_1a_2\ldots a_8],b}$ and $R^{[a_1a_2\ldots a_8,b
]}=0$. The indices $a,b,\ldots =1,2\ldots 11$.  The generators $K^a{}_b$ are those of GL(11) and have level zero, the other generators have levels $1,2,3,.....$. 
The  negative level generators  are  given by 
$$ 
R_{a_1a_2a_3}, \ R_{a_1a_2\dots a_6} , \ R_{a_1a_2\ldots a_8,b},\ldots
\eqno(5.2)$$
where the last generator obey an analogous constraint. 
\par
The $E_{11}$ algebra can be constructed using the Serre procedure but it is much easier to construct it level by level using the known generators at a given level, the fact that the commutators preserve the level and  obey the Jacobi identity. The GL(11) generators, by definition,  obey the commutators 
$$
[K^a{}_b , K^c{}_d ] = \delta ^b_c K^a{}_d - \delta ^a_d K^b{}_c 
\eqno(5.3)$$
The level one commutators are those between the generators of SL(11) and the generators $R^{a_1a_2a_3}$ and are given by 
$$
[K^a{}_b , R^{c_1c_2c_3} ] = \delta _b^{c_1} R^{a  c_2c_3}+\delta _b^{c_2} R^{c_1 a c_3}+\delta _b^{c_3} R^{c_1c_2 a}
\eqno(5.4)$$
It just expresses the fact that the generators belong to a representation of SL(11). 
\par
At the next level we find the commutator 
$$
[ R^{a_1a_2a_3}, R^{b_1b_2b_3} ] = 2R^{a_1a_2a_3b_1b_2b_3} 
\eqno(5.5)$$
This latter result is obvious given that $R^{a_1a_2a_3}$ has level one and so the commutator of two of them must have level two and  must be equal to the only generator at that level. The choice of factor of 2  fixes the normalisation of the level two generator in the algebra. 
The $E_{11}$ algebra is known up to level three  [23]. The reader can find a detailed account of its construction and the result in chapter 16 of reference [16]. 
\par
The first fundamental representation of $E_{11}$, denoted the $l_1$ representation is the representation with highest weight $\Lambda_1$ which obeys the equation $(\Lambda_1, \alpha_a)=\delta_{1,a}$. We will also refer to  this representation the vector representation of $E_{11}$. As for the $E_{11}$ algebra we also consider the representation when decomposed into representations of SL(11). It can be constructed using the standard techniques involving  raising and lowing generators.  One finds that the vector representations contains the elements [24]
$$
P_a, Z^{ab}, \ Z^{a_1\ldots a_5}, \ Z^{a_1\ldots a_7,b},\  Z^{a_1\ldots a_8},\ 
\  Z^{ b_1 b_2  b_3,  a_1\ldots  a_8},\ 
\  Z^{( c  d ),  a_1\ldots  a_9},\ 
$$
$$
\  Z^{ c d, a_1\ldots  a_9},\ 
\  Z^{ c, a_1\ldots  a_{10}}\ (2),\ 
Z^{a_1\ldots a_{11}} ,\ 
Z^{ c,  d_1\ldots  d_4, a_1\ldots  a_9},\ 
Z^{ c_1\ldots  c_6, a_1\ldots  a_8},\ 
Z^{ c_1\ldots  c_5, a_1\ldots  a_9},\ 
$$
$$
Z^{ d_1, c_1  c_2  c_3, a_1\ldots  a_{10}},\ (2),\  Z^{ c_1 \ldots  c_4, a_1\ldots  a_{10}},\ (2),\  Z^{( c_1 c_2, c_3 ),a_1\ldots a_{11}},\ 
Z^{ c, a_1 a_2},\ (2),\  
Z^{ c_1\ldots  c_{3},a_1\ldots a_{11}},\ (3),\ 
 \ldots  
\eqno(5.6)$$
The blocks of indices contain indices that are totally antisymmetrised while  $()$ indicates that the indices are symmetrised. All the elements come with multiplicity one except when there is a bracket which gives the multiplicity. All the generator belong to irreducible representations of SL(11), for example $Z^{a_1\ldots a_7,b}$ obeys the constraint $Z^{[ a_1\ldots a_7,b ]}=0$. The $l_1$ generators  are also classified by a level which is the number of up minus down indices plus one. 
 We have listed the generators   up to and including level five. 
\par
The level zero entry  follows from the observation that, at level zero,  we delete node eleven in the $E_{11}$ Dynkin diagram leaving the SL(11) algebra and so we have the first fundamental representation of SL(11) which is a vector of SL(11), that is, $P_a$. We note that the first three entries have the same form as the central charges of the eleven dimensional supersymmetry algebra, but these are only a small part of the vector representation. In fact $E_{11}$ seems to systematically predict results which are usually considered to follow from supersymmetry. 
\par
The charges for the point particle,  the two brane and five brane are the first three objects respectively in the $l_1$ representation. There is very good evidence that the $l_1$ representation contains all branes charges [24,25,26,27]
\par
To construct the algebra $E_{11}\otimes_s l_1$ we promote the elements  of the vector representation to be generators and then find the commutators of equation (2.3) for this case. The simplest way is to proceed level by level preserving the level and implementing the Jacobi identity. One finds for the first two levels that [24] 
$$
[K^a{}_b , P_c ]= - \delta ^a_c P_b +{1\over 2} \delta^a_b P_c
\eqno(5.7)$$
$$
[R^{a_1a_2a_3} , P_b ]= 3\delta _b^{[a_1} Z^{a_2a_3]}
\eqno(5.8)$$
We take the $l_1$ generators to commute. 
The $E_{11}\otimes_s l_1$ algebra is known  up to level four and  can be found up to level three in chapter 16  of [16] where a detailed account of this algebra and its construction can be found. This includes the perhaps unexpected  extra term in equation (5.7). 
\par
The lists of $E_{11}$ and $l_1$ generators to quite high levels can be found   in reference [16] and by using the Nutma programme Simplie [28]. 

\medskip
{\bf 6. The non-linear realisation of $E_{11}\otimes _s l_1$ }
\medskip
To construct this non-linear realisation $E_{11}\otimes _s l_1$ we follow the procedure given in section two for a non-linear realisation of type III. We need to know not only the algebra $E_{11}\otimes _s l_1$, discussed in the last section,  but also the local subalgebra. 
Any Kac-Moody algebra possess an involution, called the Cartan involution. This  takes the generators associated with positive roots into generators associated with the corresponding negative roots. We take the local subalgebra of the non-linear realisation $E_{11}\otimes _s l_1$ to be the subalgebra of $E_{11}$  that is invariant under this Cartan involution. We denote this algebra by $I_c(E_{11})$ and it contains the generators 
$$
I_c(E_{11})= \{ K^a{}_b - \eta^{ac}\eta_{bd}K^d{}_{c} ,\  R^{a_1a_2a_3} - \eta ^{a_1b_1}\eta^{a_2b_2}\eta^{a_3b_3}R_{b_1b_2b_3}, \ldots \}
\eqno(6.1)$$
We notice that it does, as expected, involve generators that are a sum of positive and negative level generators. The level zero generators of $I_c(E_{11})$ are the first entry in equation (6.1), are those of SO(11), which is consistent with the fact that the  Cartan involution invariant subalgebra of SL(D) is SO(D).  The Cartan involution invariant subalgebras for the real forms of the algebras in the Cartan-Killing list are  the maximally compact algebras,  for example, the  Cartan involution invariant subalgebra of  $E_7$  is $I_c(E_7)=SU(8)$ and for $E_8$ it is $I_c(E_8)=SO(16)$.  
A detailed discussion of the  algebra $I_c(E_{11})$ and the more complete mathematical  definition of the Cartan involution  for any Kac-Moody algebra can be found in chapter sixteen of reference [16]. 
\par
The classification, and many of the properties of Kac-Moody algebras, are   usually investigated by taking  the algebras to be  over the complex numbers. However, in the application we have in mind we will use a particular real form. It is simpler to take the real form such that we find SL(11) and then do a Wick rotation at the end of the calculation to find SO(1,10) in $I_c(E_{11})$. However, one can also take a variant of the usual Cartan involution such that the Cartan involution invariant subalgebra $I_c(E_{11})$ contains  SO(1,10) [26] and the work with the Lorentz group from the very beginning. We largely, but not always,  follow the former path. 
\par
The non-linear realisation of $E_{11}\otimes _s l_1$ with local subalgebra 
$I_c(E_{11})$ is constructed from a  group element of $E_{11}\otimes_s l_1$ which can be written in the form $g=g_l g_E$ where [23, 24]
$$
g_E=  \ldots e^{ h_{a_1\ldots a_{8},b}
R^{a_1\ldots a_{8},b}} e^{ A_{a_1\ldots
a_6} R^{a_1\ldots a_6}}e^{ A_{a_1\ldots a_3} R^{a_1\ldots 
a_3}} e^{h_a{}^b K^a{}_b} \equiv e^{A_{\underline \alpha} R^{\underline \alpha}}
\eqno(6.2)$$ 
and $$
g_l= e^{x^aP_a} e^{x_{ab}Z^{ab}} e^{x_{a_1\ldots a_5}Z^{a_1\ldots a_5}}\ldots \equiv
e^{z^A L_A} 
\eqno(6.3)$$
We have used the local subalgebra $I_c(E_{11})$ of equation (6.1)  to choose the group element $g_E$ to have no negative level generators. In doing this we used all of the local symmetry except for the local Lorentz transformation which remain a local symmetry. Apart form the level zero generators, the group element $g_E$  lies in the Borel subalgebra of $E_{11}$. 
\par
The corresponding theory will contain the fields [23]
$$
h_a{}^b, \ A_{a_1\ldots a_3}, \ A_{a_1\ldots a_6}, \ h_{a_1\ldots a_{8},b},\ldots 
\eqno(6.4)$$
which depend on a spacetime that has the coordinates [24]
$$
x^a , \ x_{ab} ,\ x_{a_1\ldots a_5},,\ x_{a_1\ldots a_8},\ x_{a_1\ldots a_7b},\ldots 
\eqno(6.5)$$
The next few higher level coordinates can be read off from equation (5.6).
\par
Thus one finds at level zero and one the fields of the usual formulation of eleven dimensional supergravity theory; the graviton and the three form. The  field at level two is  a six form which is well known to provide an equivalent description of the degrees of freedom that are usually carried by the three form. The field at level three 
$h_{a_1\ldots a_{8},b}$,  provides a dual description of gravity;  indeed it was in [23] that the linearised equation of motion, that was first order in spacetime derivatives and  that express  this duality were  formulated for the field $h_{a_1\ldots a_{D-3}, b}$ in any dimension. These equations guaranteed that the dual fields really did describe gravity at the linearised level. 
However, above level three the non-linear realisation contains an infinite number of fields. The  physical role of these higher level fields was unfamiliar to us at the early stages of  work on $E_{11}$, but we now understand this role for quite large classes of the fields. We will discuss this later in the review but this still leaves  many fields whose role is unknown. 
\par
The spacetime possess the usual coordinates of eleven dimensional spacetime, but it also has  many more coordinates, in fact it is an infinite dimensional spacetime.  An initially  somewhat intimidating prospect. It can be shown that for every element in the Borel subgroup of $E_{11}$ there is at least one corresponding element in the $l_1$  representation [25].  For example $K^a{}_b$ and $R^{a_1a_2a_3}$ correspond to $P_a$ and $Z^{a_1a_2}$ respectively; the general pattern being that one just knocks an index off the generators in the Borel subalgebra of $E_{11}$. However in the non-linear realisation every element in the Borel subalgebra of $E_{11}$ leads to a field and every element in the $l_1$ representation leads to a coordinate of the spacetime. As a result we find that every field leads to at least one coordinate, a fact that is evident at low levels and is given in the correspondence below. 
$$
\matrix { x^a \leftrightarrow P_a \leftrightarrow K^a{}_b \leftrightarrow h_a{}^b\cr
x_{a_1a_2}\leftrightarrow Z^{a_1a_2}\leftrightarrow R^{a_1a_2a_3} \leftrightarrow A_{a_1a_2a_3}\cr
x_{a_1\ldots a_5}\leftrightarrow Z^{a_1\ldots a_5}\leftrightarrow R^{a_1\ldots a_6} \leftrightarrow A_{a_1\ldots a_6}\cr
x_{a_1\ldots a_8} ,\ x_{a_1\ldots a_7,b}\leftrightarrow Z^{a_1\ldots a_8} ,\ Z^{a_1\ldots a_7,b} \leftrightarrow R^{a_1\ldots a_8,b} \leftrightarrow h_{a_1\ldots a_8,b}\cr}
$$
\par
We see from the above correspondence  that the graviton is associated with the usual coordinates of spacetime $x^a$ which carries the effects of gravity through the curvature of spacetime. The three form field is associated with the two form coordinates, the six form with the five form coordinate and so on.  What this implies  is that the $E_{11}$ symmetry which rotates the graviton into the three form and the higher fields also requires an extension of our notion of spacetime with a corresponding new geometry associated with the new fields beyond those of gravity. In this context it is interesting to recall the following quote taken from Salam's nobel lecture [30];

"...But are all the fundamental forces gauge forces? Can they be understood as such, in terms of charges- and their corresponding currents-only? And if they are how many charges? What unified entity are the charges components of? what is the nature of charge? Just as Einstein comprehended the nature of the gravitational charge in terms of space-time curvature, can we comprehend the nature of other charges-the nature of the entire unified set, as a set, in terms of something equally profound? This briefly is  the dream..."
\par
We now outline how to construct the dynamics of the $E_{11}\otimes _s l_1$ non-linear realisation.   The Cartan forms were defined  in section two in equations (2.13-15) for the case of a general  type III non-linear realisation and  in this case  they can be written as 
$$
{\cal V} = dx^\Pi E_\Pi{}^A l_A + dx^\Pi G_{\Pi , } {\underline \alpha} R^{\underline \alpha} 
\eqno(6.6)$$
where $G_{\Pi , } {\underline \alpha}$ are the Cartan forms of $E_{11}$. As previously noted we now denote the generators of $E_{11}$ by $R^{\underline\alpha}$; the use of the underline being required to avoid ambiguity with the index $\alpha$ that arises in the discussion of the supergravities theories in lower dimensions (see next section). 
\par
The  transformations of the Cartan forms under the symmetries of the non-linear realisation were given in equations (2.15) and (2.16).   
Although the Cartan forms, when  viewed as forms, are   inert under rigid transformations,  the rigid transformations do  act on
the coordinate differentials, that is, on the $dx^\Pi$,  contained in the Cartan form. This action  induces  a corresponding  rigid $E_{11}$ transformation   on the lower index of ${ E_\Pi{}^A}$. 
Indeed, it follows from equation (2.11) that the coordinates are inert under the local transformations but transform under the rigid  transformations as 
$$
z^A l_A\to z^{A\prime} l_A=g_0 z^Al_A g_0^{-1} = z^\Pi D(g_0^{-1})_\Pi {}^Al_A
\eqno(6.7)$$
When  written  in matrix form the differential  transformations act as  
$dz^T
\to dz^{T\prime}= dz ^T D(g_0^{-1})$.   As a result the derivative
$\partial_\Pi\equiv {\partial\over \partial z^\Pi}$ in the generalised space-time 
transforms as $\partial_\Pi^\prime= D(g_0)_\Pi{}^\Lambda \partial_\Lambda$. 
\par 
A local $I_c(E_{11})$ transformation acts on the $\underline \alpha$ index of $G_{\Pi, \underline \alpha}$ and on the $A$ index of ${E}_\Pi{}^{A} $ as governed by  equation (2.16).    As a result the rigid and local transformations of the object ${ E_\Pi{}^A}$
can  be summarise  as 
$$
{ E}_\Pi{}^{A\prime} =
D(g_0)_\Pi{}^\Lambda { E}_\Lambda{}^{B}D(h)_B{}^A 
\eqno(6.8)$$
and  for  its   inverse by  
$$ (E^{-1})_A{}^{\Pi\prime}= D(h^{-1})_A{}^B (E^{-1})_B{}^\Lambda 
D(g_0^{-1})_\Lambda{}^\Pi
\eqno(6.9)$$
where $h^{-1} l_A h \equiv D(h)_A{}^B l_B$. 
Thus  the object  $E_\Pi{}^A$ transforms
under a local $I_c(E_{11})$ transformation on its $A$ index and by a rigid $E_{11}$ induced 
coordinate transformation of the generalised space-time on its $\Pi$ index.  These transformations mean that we can interpret  ${ E}_\Pi{}^{A}$ as a vielbein of the  space-time which possess the tangent group $I_c(E_{11})$. As we noted above, at level zero  $I_c(E_{11})$ is just SO(1,10). The reader may be puzzled by the use of the indices $\Pi, \Lambda , \ldots $ rather than $A,B,\ldots $ to label the elements  of the $l_1$ representation, but this just reflects whether the indices transform under the rigid, or local transformations, that is, are world or tangent indices respectively. 
\par
Similarly, the object $G_{\Pi , }{}_{\underline \alpha}$ transforms by  the same a rigid $E_{11}$ induced 
coordinate transformation  on its $\Pi$ index and by  under a local $I_c(E_{11})$ transformation on its $\underline \alpha$ index. As a result we find that 
the object $G_{A , } {}_{\underline \alpha} \equiv (E^{-1})_A{}^\Pi G_{\Pi , }{}_{\underline \alpha}$ is inert under the rigid 
$E_{11}\otimes _s l_1$ transformations and only transforms under the 
$I_c(E_{11})$ transformations. To find the dynamics we  can use the objects $G_{A , } {}_{\underline \alpha}\  $ ,  in this case the equations will be automatically invariant under the rigid transformations and we only need to solve the problem of finding  a set of equations which is invariant under $I_c(E_{11})$ transformations. 
\par
It is very straightforward to compute the vielbein using the form of the group element of equation (6.2) and its  definition  as given in equation (6.6). One finds that the vielbein up to level two is given by [31,32]
$$
{ E}= (\det e)^{-{1\over 2}}
\left(\matrix {e_\mu{}^a&-3 e_\mu{}^c A_{cb_1b_2}& 3 e_\mu{}^c A_{cb_1\ldots b_5}+{3\over 2} e_\mu{}^c A_{[b_1b_2b_3}A_{|c|b_4b_5]}\cr
0&(e^{-1})_{[b_1}{}^{\mu_1} (e^{-1})_{b_2]}{}^{\mu_2}&- A_{[b_1b_2b_3 }(e^{-1})_{b_4}{}^{\mu_1} (e^{-1})_{b_5 ]}{}^{\mu_2}  \cr
0&0& (e^{-1})_{[b_1}{}^{\mu_1} \ldots (e^{-1})_{b_5]}{}^{\mu_5}\cr}\right)
\eqno(6.10)$$
\par
Using the form of the group element of equation (6.2),  the Cartan forms of equation (6.6) can  readily by found to be given, up to level three,  by [23,31]
$$
G _{a}{}^b=(e^{-1}d e)_a{}^b,\ \ G_{a_1\ldots a_3}= e_{a_1}{}^{\mu_1}\ldots e_{a_3}{}^{\mu_3}
dA_{\mu_1\ldots \mu_3}, 
\eqno(6.11)$$
$$ 
G_{a_1\ldots a_6}=  e_{a_1}{}^{\mu_1}\ldots e_{a_6}{}^{\mu_6}(d A_{\mu_1\ldots \mu_6} 
- A_{[ \mu_1\ldots \mu_3}d A_{\mu_4\ldots \mu_6]})
\eqno(6.12)$$
where we are writing the quantities as forms. 
\par
In order to better understand some of the early $E_{11}$ papers it  is instructive to recall their progress  towards constructing the non-linear realisation. Taking reference [22] and using [23] together one finds that the eleven dimensional supergravity was constructed, at very low levels,  as   a non-linear realisation of $E_{11}$; it was shown to be a non-linear realisation of a particular algebra  in reference [22] and the generators in this algebra are identified as those of $E_{11}$ in reference  [23]. However, this calculation suffered from a number of shortcomings.  It only introduced the usual spacetime translation generators,  which was not a $E_{11}$ covariant procedure,  and consequently the resulting field theory  only  possessed the usual spacetime. Also it only enforced the $I_c(E_{11})$ symmetry at the lowest level, that is, the very  weak Lorentz part. As a result,  the non-linear realisation carried out with these limitations did not lead uniquely to eleven dimensional supergravity and one had to fix several constants whose values were not determined by the calculation. A more systematic approach was taken in references [31,48, 33] and [34] where the non-linear realisation of $E_{11}\otimes_s l_1$ at low levels was constructed for the fields up to an including the dual graviton as well as  the low level coordinates of the $l_1$ representation. These references  enforced not only the  Lorentz group symmetries of $I_c(E_{11})$ but also  the much more powerful symmetries at the next levels.  The  approach of reference [31] focused on finding  duality equations which were first order in derivatives and it  found the correct equations for the forms which were uniquely determined but there were unresolved issues with the graviton sector. In references [33, 34] the invariant second order equations were found, they were unique and when one retained only the low levels fields and the level zero coordinates they were precisely those of eleven dimensional supergravity. This essentially proved the $E_{11}$ conjecture. 
\par
We recall that  the Cartan involution invariant subalgebra $I_c(E_{11})$ at lowest level  is SO(1,10). At the next level $I_c(E_{11})$ possess a group element  $h$ which  involves the 
 generators at levels $\pm 1$ and it  is of the form [31]
$$
h=1- \Lambda _{a_1 a_2  a_3 }S^{a_1 a_2  a_3  }, \quad {\rm  where   }\quad
S^{a_1a_2a_3}= R^{a_1a_2a_3}- \eta^{a_1b_1} \eta^{a_2b_2}\ \eta^{a_3b_3} R_{b_1b_2b_3}
\eqno(6.13)$$
Under this transformation the Cartan forms of equation (6.6), when written as forms, change as     
$$
\delta\,{\cal V}_E = \left[S^{a_1a_2a_3}\,\Lambda_{a_1a_2a_3},{\cal V}_E\right] - S^{a_1a_2a_3}\,d\Lambda_{a_1a_2a_3}.
\eqno(6.14)$$ 
The local  $I_c(E_{11})$ variations of the Cartan forms are straightforward to compute, using the $E_{11}$ algebra  and they are given by  [31,33,34]
$$
\delta G_{a}{}^{b}=18 \Lambda^{c_1c_2 b }G_{c_1c_2 a}
-2 \delta_a ^{b}  \Lambda^{c_1c_2 c_3}G_{c_1c_2 c_3},\ 
\eqno(6.15)$$
$$
\delta G_{a_1a_2a_3}=-{5!\over 2} G_{b_1b_2b_3 a_1a_2a_3}
\Lambda^{b_1b_2 b_3}-3G^{c}{}_{[a_1 } \Lambda_{|c|a_2a_3]} -d \Lambda_{a_1a_2a_3}
\eqno(6.16)$$
$$
\delta G_{a_1\ldots a_6}=2 \Lambda_{[ a_1a_2a_3}G_{a_4a_5a_6 ]}
-8.7.2 G_{b_1b_2b_3 [ a_1\ldots a_5,a_6]}\Lambda^{b_1b_2b_3}
+8.7.2 G_{b_1b_2[ a_1\ldots a_5a_6, b_3 ]}\Lambda^{b_1b_2b_3}
\eqno(6.17)$$
\par
The above transformations do not take account of the fact that the  $l_1$ index on the Cartan forms can transform. As explained above if this index is made into a tangent index, that is, $G_{A , \underline \alpha} = (E^{-1})_A{}^\Pi G_{\Pi , \underline \alpha}$ it  transforms only under the local $I_c(E_{11})$ transformations, the transformation  just being that for the inverse vielbein of equation (6.8). One finds that the Cartan forms, when referred to the tangent space,  transforms  on their $l_1$ index as [31,33,34]
$$
\delta G_{a, \bullet}= -3G^{b_1b_2}{}_{,\bullet}\ 
\Lambda_{b_1b_2 a},
\quad \delta G^{a_1a_2}{}_{, \bullet}= 6\Lambda^{a_1a_2
b}  G_{b,}{}_{\bullet} , \ldots
\eqno(6.18)$$
Of course to get the full transformation one must combine the transformations of equation (6.18) with those of equations (6.12-6.17); for example we find that 
$$
\delta G^{e_1e_2}{}_{, a}{}^{b}=18 \Lambda^{c_1c_2 b }G^{e_1e_2}{}_{,}{}_{c_1c_2 a}
-2 \delta_a ^{b}  \Lambda^{c_1c_2 c_3}G^{e_1e_2}{}_{,}{}_{c_1c_2 c_3}
+6\Lambda^{e_1e_2d} G_{d}{}_{, a}{}^{b} 
\eqno(6.19)$$
\par
The detailed construction of the equations of motion which follow from the 
$E_{11}\otimes_s l_1$ non-linear realisation was given in reference [34] following earlier results in references [33] and [31]. We refer the reader to this reference and confine ourselves here to  stating the result. One finds the unique equations of motion are given by 
$$
{\cal E} ^ {a_1a_2a_3}\equiv {1\over 2}  G_{b,d}{}^{d} G^{[b, a_1a_2a_3]}- 3G_{b,d}{}^{[a_1|} G^{[b, d |a_2a_3]]}
-G_{c,b}{}^{c} G^{[b, a_1a_2a_3]}+ (\det e)^{{1\over 2}}e_b{}^\mu\partial_\mu G^{[b, a_1a_2a_3]}
$$
$$
+{1\over 2.4!}\epsilon ^{a_1a_2a_3b_1\ldots b_8} G_{[b_1,b_2 b_3 b_4]} G_{[b_5, b_6 b_7b_8] }
$$
$$  -\,9\,G^{ca_1}{}_{,cd_1d_2}\,G^{[d_1,\,d_2a_2a_3]} + {5\over 16}\,\varepsilon^{a_1a_2a_3b_1...b_8}\,G_{b_1,\,b_2b_3b_4}\,G^{c_1c_2}{}_{,c_1c_2b_5...b_8}
$$
$$
+ {1\over 4}\,e_{\mu_1}{}^{[a_1}e_{\mu_2}{}^{a_2}e_{\mu_3}{}^{a_3 ]}
\partial_\nu \left( (\det e)^{{1\over 2}} G^{\mu_1\mu_2}{}_{,}{}^{\nu\mu_3 }\right)
+{1\over 4} (\det e)^{{1\over 2}} \omega_{\nu,} {}^{[a_1 |b} G^{a_2a_3 ]}{}_{, b}{}^{\nu}
$$
$$
+{1\over 4} G^{[a_1a_2 |}{}_{, d} {}^{d} (G^{|a_3 ] }{}_{, c}{}^{c}-G_{c,}{}^{ |a_3 ]c})
-{1\over 4} \partial_\nu \left( (\det e)^{{1\over 2}}  (G^{[ a_1a_2 |}{}_{,d}{}^{d}e^{\nu |a_3]}-
G^{[ a_1a_2 |}{}_{,}{}^{ |a_3 ]\nu} )\right)
$$
$$
+{1\over 2} ( G^{[a_1}{}_{,c}{}^{a_2} G^{|c | a_3]}{}_{,d}{}^{d} -G^{c [a_1}{}_{,} {}^{a_2 |e|} G_{e,c}{}^{a_3]})
$$
$$
+{15\over 2} e_{\mu_1}{}^{[a_1}e_{\mu_2}{}^{a_2}e_{\mu_3}{}^{a_3 ]}
\partial_\nu\left((\det e)^{{1\over 2}} G^{d_1d_2}{}_{, d_1d_2} {}^ {\nu \mu_1\mu_2\mu_3}\right)
$$
$$
+e_{\mu_1}{}^{[a_1}\,e_{\mu_2}{}^{a_2}\,e_{\mu_3}{}^{a_3]}\,\Big({1\over 2}\,\left(det\,e\right)^{1\over 2}\,g_{\tau\sigma}\,g^{\mu_1\lambda}\,\partial_{\lambda}\,G^{\tau\mu_2,\,\left(\sigma\mu_3\right)} - {1\over 2}\,G^{\tau\mu_1}{}_{,\,d}{}^d\,G^{\mu_2,\,(\mu_3}{}_{\tau)} 
$$
$$
- {1\over 4}\,G^{\tau\mu_1,\,(\mu_2}{}_{\tau)}\,G^{\mu_3}{}_{,\,d}{}^d-\,G^{\tau\mu_1}{}_{,\,(\tau\sigma)}\,G^{\mu_2,\,(\mu_3\sigma)} + G^{\tau\mu_1,\,(\mu_2\sigma)}\,G_{\sigma,\,(\tau}{}^{\mu_3)}\Big)=0
\eqno(6.20)$$
and 
$$
{\cal E}_{a b}\equiv (\det e)  {\cal R}_{ab}- 12.4 G_{[a, c_1c_2c_3]}G^{[e, c_1c_2c_3]}\eta_{eb}+4\eta_{ ab} G_{[c_1, c_2c_3c_4]}G^{[c_1, c_2c_3c_4]}
$$
$$
-\,3.5!\,G^{d_1d_2}{}_{,\,d_1d_2a}{}^{c_1c_2c_3}\,G_{[b,\,c_1c_2c_3]} - 3.5!\,G^{d_1d_2}{}_{,\,d_1d_2b}{}^{c_1c_2c_3}\,G_{[a,\,c_1c_2c_3]} 
$$
$$+ {5!\over 2}\,\eta_{ab}\,G^{d_1d_2}{}_{,\,d_1d_2c_1...c_4}\,G^{[c_1,\,c_2c_3c_4]}
 - 12\,G^{c_1c_2}{}_{,\,a}{}^{c_3}\,G_{[b,\,c_1c_2c_3]} 
+3 G^{c_1c_2}{}_{, e}{}^{e} G_{[a,bc_1c_2]}
$$
$$
-6\left(\det\,e\right)\,e_{b}{}^\mu\,e_a{}^\lambda\,\partial_{[\mu |}\left[\left(det\,e\right)^{-\,{1\over 2}}G^{\tau_1\tau_2}{}_{,\,| \lambda\tau_1\tau_2]}\right]
$$
$$
-\,\left(det\,e\right)^{{1\over 2}}\omega_c{}_{,}{}_{b}{}^{c}\,G^{d_1d_2}{}_{,\,d_1d_2a} - 3\,\left(det\,e\right)^{{1\over 2}}\omega_a{}_{,}{}_{b}{}^{c}\,G^{d_1d_2}{}_{,\,d_1d_2c}
=0
\eqno(6.21)$$
where 
$$
{\cal R}_a{}^b = e_a{}^\mu \partial_\mu\,\Omega_{\nu,}{}^{bd}\,e_d{}^\nu - e_a{}^\mu \partial_\nu\,\Omega_{\mu,}{}^{bd}\,e_d{}^\nu + \Omega_{a,}{}^b{}_c\,\Omega_{d,}{}^{cd} - \Omega_{d,}{}^b{}_c\,\Omega_{a,}{}^{cd} , 
\eqno(6.22)$$
$$
(\det e)^{{1\over 2}} \omega _{c, ab}= - G_{a, (bc)}+ G_{b, (ac)}+G_{c, [ab]}
\eqno(6.23)$$
and 
$$
(\det e)^{{1\over 2}}\Omega _{c, ab}= (\det e)^{{1\over 2}}\omega_{c, ab}
-3 G^{dc}{}_{, dab} -3 G^{d}{}_{b}{}_{, dac} +3 G^{d}{}_{a}{}_{, dbc} 
-\eta _{bc} G^{d_1d_2}{}_{, d_1d_2 a}+\eta _{ac} G^{d_1d_2}{}_{, d_1d_2 b}
\eqno(6.24)$$
We have corrected the sign of the eleventh term compared to that in reference [34]. 
\par
Under the $I_c(E_{11})$ transformation of equation (6.13) they transform as 
$$
\delta {\cal E}^{a_1a_2a_3 }= {3\over 2} E_b{}^{[a_1|}\Lambda ^{b | a_2a_3]} 
$$
$$
+{1\over 24} e_{\mu_1}^{a_1}e_{\mu_2}^{a_2} e_{\mu_3}^{a_3} \epsilon ^{\mu_1\mu_2\mu_3 \nu\lambda_1\ldots \lambda_4 \tau_1\tau_2\tau_3}\partial_\nu \left( (\det e)^{-{1\over 2}} E_{\lambda_1\ldots \lambda_4}g_{\tau_1 \kappa_1} g_{\tau_2 \kappa_2}
g_{\tau_3 \kappa_3}\right) \Lambda^{ \kappa_1\kappa_2\kappa_3}
$$
$$
+{1\over 24.4!} \epsilon^{a_1a_2a_3b_1\ldots b_8} \epsilon _{b_1\ldots b_4 c_1c_2c_3 e_1\ldots e_4} E_{b_5\ldots b_8} G^{[e_1,e_2\ldots e_4 ]} \Lambda ^{c_1c_2c_3}
\eqno(6.25)$$
and 
$$
\delta {\cal E}_{ab}= -36 \Lambda ^{d_1d_2 }{}_{a} { E}_{bd_1d_2} -36 \Lambda ^{d_1d_2 }{}_{b} { E}_{ad_1d_2} 
+8\eta_{ab}\Lambda ^{d_1d_2d_3 } E_{ d_1d_2d_3} 
$$
$$
-2 \epsilon_{a}{}^{c_1c_2c_3 e_1\ldots e_4 f_1f_2f_3} \Lambda_{f_1f_2f_3} 
E_{e_1\ldots e_4}G_{[b, c_1c_2c_3]}
-2 \epsilon_{b}{}^{c_1c_2c_3 e_1\ldots e_4 f_1f_2f_3} \Lambda_{f_1f_2f_3} 
E_{e_1\ldots e_4}G_{[a, c_1c_2c_3]}
$$
$$ +{1\over 3} \eta_{ab} \epsilon^{c_1\dots c_4 e_1\ldots e_4 f_1f_2f_3}
E_{e_1\ldots e_4}G_{[c_1, c_2c_3c_4]}\Lambda_{f_1f_2f_3} 
\eqno(6.26)$$
In these variations 
$$
E_{a_1\ldots a_4}\equiv {\cal G}_{[a_1,a_2a_3a_4] }-{1\over 2.4!}\epsilon _{a_1a_2a_3a_4}{}^{b_1\ldots b_7} G_{b_1,b_2\ldots b_7 }=0
\eqno(6.27)$$
where 
$$
{\cal G}_{a_1,a_2a_3a_4 }\equiv  G_{[a_1,a_2a_3a_4] }+{15\over 2}G^{b_1b_2}{}_{, b_1b_2 a_1\ldots a_4}
\eqno(6.28)$$
This is the first order duality relation between the three form and six form fields. It can be found independently by requiring an equation which is first order in derivatives and contains the three form and is part of an invariant set of equations under  the transformations of  (6.13) [31,34]. When carrying out the  variation of this equation one also finds the duality relation between the usual graviton and the dual graviton. 
We note that taking the spacetime derivative we find, at least, to lowest orders the second order equation of motion of equation (6.20). 
\par
If we discard the derivatives with respect to the higher level coordinates we find that the above equations of motion  can be written as 
$$
\partial_{\nu}( (\det e)^{{1\over 2}} G^{[\nu,\mu_1\mu_2\mu_3]})+
{1\over 2.4!} (\det e)^{{-1}}\epsilon ^{\mu_1\mu_2\mu_3\tau_1\ldots\tau_8} G_{[\tau_1,\tau_2\tau_3\tau_4]} G_{[\tau_5,\tau_6\tau_7\tau_8] }=0 
\eqno(6.29)$$
and that 
$$
E_a{}^b\equiv (\det e)  R_a{}^b- 12.4 G_{[a, c_1c_2c_3]}G^{[b, c_1c_2c_3]}+4\delta _a^b G_{[c_1, c_2c_3c_4]}G^{[c_1, c_2c_3c_4]}=0
\eqno(6.30)$$
We recognise these are the equations of motion of eleven dimensional supergravity. Thus the $E_{11}\otimes_s l_1$ non-linear realisation lead to unique equations, at least up to the levels studied, and when truncated to contain the low level fields and only the usual coordinates of spacetime these equations of motion these equations of motion are precisely those  of the bosonic sector of eleven dimensional supergravity. The reader who repeats even parts of these calculations will be left in no doubt as the validity of the $E_{11}$ approach. 
\medskip 
{\bf 7 The $E_{11}\otimes_s l_1$ non-linear realisation in $D$ dimensions}
\medskip
In the section five  we considered the $E_{11}\otimes_s l_1$  algebra when decomposed with respect to its  GL(11) subalgebra and we found that the $E_{11}\otimes_s l_1$ non-linear realisation, when decomposed in this way, was an eleven dimensional theory that at low levels was precisely eleven dimensional supergravity. To find the theory in $D$ dimensions we delete the node labelled $D$ in the $E_{11}$ Dynkin diagram to find  the residual algebra  
$GL(D)\otimes E_{11-D}$, we then decompose  the $E_{11}\otimes_s l_1$ algebra into representations of this subalgebra and then construct the corresponding non-linear realisation [35,36,37,38]. 
\medskip
$$
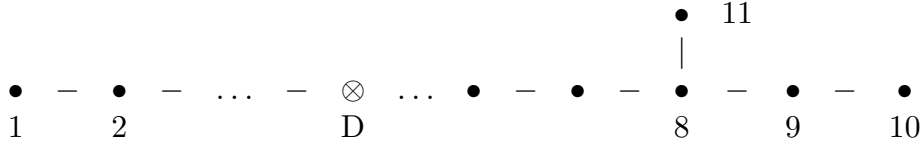
\matrix{
& & & & & & & & & & & &  \bullet & 11 & & & \cr 
& & & & & & & & & & & & | & & & & \cr
\bullet & - & \bullet & - & \ldots  & - & \otimes  & \ldots  & \bullet & - & \bullet & - & \bullet &-&\bullet&-&\bullet&\cr
1 & & 2 & &  & & $D$ &   & &  &  & & 8 & & 9 & & 10 \cr
}
$$
\medskip
In this non-linear realisation the $GL(D)$ subalgebra will lead  to gravity in $D$ dimensions,  confirming the fact that the resulting  theory is indeed in $D$ dimensions. The $E_{11-D}$ subalgebra is the well known U duality algebra of the supergravity theory in $D$ dimensions. 
\par
Carrying out the decomposition we find at low levels exactly the fields of the $D$ dimensional maximal supergravity theory and a generalised spacetime whose level zero part is just the usual spacetime in $D$ dimensions. For example,  in five dimensions one deletes node five to find the remaining algebra $GL(5)\otimes E_6$ and decomposing with respect to this subalgebra one finds the resulting non-linear realisation has the field content [38,33]
$$
h_a{}^b, \ \varphi_{\alpha}, \ A_{aM}, \ A_{a_1a_2}{}^{N}, \ A_{a_1a_2a_3,\,\alpha}, \ A_{a_1a_2,\,b}, \ \ldots
\eqno(7.1)$$
and the spacetime has the coordinates [38,33]
$$
x^a, \ x_{N}, \ x_{a}{}^{N}, \ x_{a_1a_2,\alpha}, \ x_{ab}, \ \ldots
\eqno(7.2)$$ 
For all these objects the lower (upper) case indexes $a,\,b,\,c,\,... = 1,\,...,\,5$ correspond to 5 ($\bar 5$)-dimensional fundamental representation of $GL(5)$ . The  indexes $\alpha,\,\beta,\,\gamma,\,... = 1,\,...,\,78$ correspond to 78-dimensional adjoint representation of $E_6$ and the 
upper and lower case  indexes $N,\,M,\,P,\,... = 1,\,...,\,27$ correspond to ${\overline 27}$-dimensional and ${27}$-dimensional representations of $E_6$  respectively. 
\par
As in eleven dimensions the fields and coordinates are classified  by a level. However the definition of the level depends on the node being deleted, we refer the reader to  reference [16] for a detailed account. For theories in less than ten dimensions the level of the fields  is just the number of lower  minus upper GL(D) indices. While for the coordinates it is the same but minus one. The fields in five dimensions, given in  equation (7.1),  are the graviton and  the scalars at level zero, while at level one we find  the vectors. Thus we find the bosonic fields of  the usual description of five-dimensional supergravity. The level two fields provide a dual description of the vectors and the two fields at level three are  a dual description of the scalars and the graviton respectively. 
The equations of motion that follow from the $E_{11}\otimes_s l_1$ non-linear realisation leading to the five dimensional theory were found, at low levels, in reference [33]. They were the equations of motion of five dimensional maximal supergravity. In reference [33] some undetermined constants appear but they are fixed to the required values by considering the dimensional reduction of the unique eleven dimensional eqations of motion which follow from the $E_{11}\otimes_s l_1$ non-linear realisation [34]. 
\par
The reader can find an account of the $E_{11}\otimes_s l_1$   non-linear realisation in the decompostion that leads to four dimensions in reference [48] where the equations of motion for the form fields are derived and are found to agree with those of maximal supergravity in four dimensions. The equations of motion of the gravity sector is only partially computed but the way is not clear  to apply the techniques of reference [33] and [34] and find the gravity equations. It is inevitable that it will agree with the equation of motion  of the maximal supergavity theory in four dimensions. 

\par
An exception to the above discussion is provided by ten dimensions. To find such a theory one has to find ten dimensional gravity and so a GL(10) subalgebra,  which includes an $A_9$ subalgebra whose Dynkin diagram  consists of nine dots in a row. Looking at the $E_{11}$ Dynkin diagram 
and starting from node one it is apparent that, unlike in less than ten dimensions where there is only one possibility, there are two possibilities. 
\par
The first possibility is to   delete node nine [35]
$$
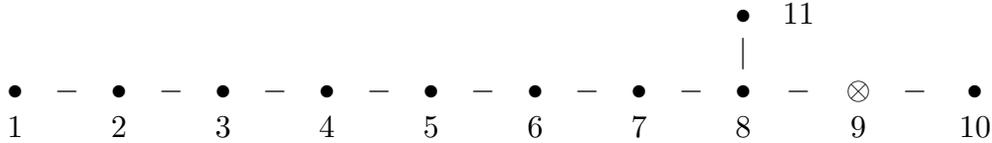
\matrix{
& & & & & & & & & & & & & & \bullet & 11 & & & \cr 
& & & & & & & & & & & & & & | & & & & \cr
\bullet & - & \bullet & - & \bullet & - & \bullet & - & \bullet & - & \bullet & - & \bullet & - & \bullet & - & \otimes & - & \bullet \cr
1 & & 2 & & 3 & & 4 & & 5 & & 6 & & 7 & & 8 & & 9 & & 10 \cr
}
$$
This  leads to the algebra $GL(10)\otimes SL(2)$. The SL(10) of the GL(10) arises from the dots one to eight as well as dot eleven. In general we refer to the  line of dots that is  is associated with gravity as the gravity line.   One finds that the field content of the resulting non-linear realisation is given by [35, 39]
$$
h_{a}{}^{b}, \ \phi, \chi ;\  A_{a_1a_2}^\alpha ;\  A_{a_1\ldots a_4} ;\  A_{a_1\ldots a_6}^\alpha ;\  A_{a_1\ldots a_8}^{(\alpha \beta)}
,\ h_{a_1\ldots a_7, b} ;\ A_{a_1\ldots a_{10}}^{(\alpha \beta\gamma)}, 
\ A_{a_1\ldots a_8, b_1 b_2}^\alpha   , 
\ A_{a_1\ldots a_9, b}^\alpha ;\ldots
\eqno(7.3)$$
where $a,b,\ldots =0,1,\ldots ,9$ are GL(10) indices and $\alpha, \beta =1,2$ are SL(2) indices. 
The first listed  fields, the graviton, the scalars, the doublet two forms and the four form  are those of the usual description of IIB supergravity. 
Node ten in the above Dynkin diagram is not connected to the gravity line and so leads to an SL(2) symmetry which is just the SL(2) symmetry  of the IIB theory. 
The $A_{a_1\ldots a_6}^\alpha $ fields are the duals of the two forms. The triplet of eight forms are the duals of the scalar fields and had previously been discussed in the  context of the IIB theory [41] but the $\bar 4$ representation of SL(2) ten forms were a prediction of $E_{11}$ [39, 40] and their presence was confirmed from the  supersymmetry perspective in [42]. The field $h_{a_1\ldots a_7, b}$ is the dual graviton.  Higher level fields can be found in Table 17.5.3 on page 596 of reference [16]. 
\par
The non-linear realisation leads to a spacetime with the coordinates 
$$ 
x^a ; \ \ x_a{}^{\alpha}  ; \ \ x_{a_1a_2a_3} ; \ \ x_{a_1\ldots a_5}{}^{\alpha} ; \ \ x_{a_1\ldots a_7} , \ \ x_{a_1\ldots a_7}{}^{(\alpha\beta)} , \ \ x_{a_1\ldots a_6, b} ; \ \ x_{a_1\ldots a_9}{}^{\alpha} (2) , \ \ x_{a_1\ldots a_9}{}^{(\alpha\beta\gamma)} , $$
$$ x_{a_1 \ldots a_8,b}{}^{\alpha} (2) , \ \ x_{a_1\ldots a_7, b_1b_2}{}^{\alpha} ; \ \ x_a (3) , \ \ x_a{}^{(\alpha\beta)} (3) , \ \ x_{a_1\ldots a_9, b_1b_2} , \ \ x_{a_1\ldots a_9, b_1b_2}{}^{(\alpha\beta)} , \ \ x_{a_1\ldots a_9, (b_1b_2)} , $$
$$ x_{a_1\ldots a_8, b_1b_2b_3}, \ \ x_{a_1\ldots a_8, b_1b_2b_3}{}^{\alpha\beta} , \ \ x_{a_1\ldots a_8, b_1b_2, c} , \ \ x_{a_1\ldots a_7,b_1\ldots b_4} \eqno(7.4) $$
where the number in brackets give the multiplicities and if there is no bracket the multiplicity is one. All the coordinates belong to irreducible representations of SL(10).  The level is the number of down minus up GL(10) indices divided by two for the fields and the same for the coordinates except that one must subtract one first.  
\par
For the second possibility, we delete node ten to find a SO(10,10) subalgebra and then delete node eleven which leads to the required GL(10) subalgebra [23]. The gravity line is made up of nodes one to nine. 
$$
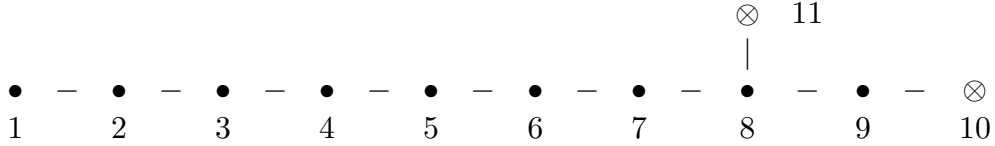
\matrix{
& & & & & & & & & & & & & & \otimes & 11 & & & \cr 
& & & & & & & & & & & & & & | & & & & \cr
\bullet & - & \bullet & - & \bullet & - & \bullet & - & \bullet & - & \bullet & - & \bullet & - & \bullet & - & \bullet & - & \otimes \cr
1 & & 2 & & 3 & & 4 & & 5 & & 6 & & 7 & & 8 & & 9 & & 10 \cr
}
$$
\par
  The corresponding field content in the non-linear realisation is given by [39]
$$
h_{a}{}^{b}(0),\ \phi (0)\ ,  A_{a_1a_2} (0);\  A_a ,\ A_{a_1a_2a_3} (1) , \  A_{a_1\ldots a_5} (1), \ A_{a_1\ldots a_7} (1), \ A_{a_1\ldots a_9} (1); 
$$
$$
A_{a_1\ldots a_6} (2),\ A_{a_1\ldots a_8} (2),\ A_{a_1\ldots a_{10}} (2),\ 
A_{a_1\ldots a_{10}} (2);\ h_{a_1\ldots a_7,b} (2),\ 
A_{a_1\ldots a_8, b_1b_2} (2), \ A_{a_1\ldots a_8, b_1b_2} (2),
$$
$$
\ A_{a_1\ldots a_9,b} (2),\ 
A_{a_1\ldots a_9,b_1b_2b_3} (2), 
;\ A_{a_1\ldots a_{10}, b_1b_2} (2), \ A_{a_1\ldots a_{10}, b_1\ldots b_4} (2);\ \ldots  
\eqno(7.5)$$
where $a,b,\ldots =0,1,\ldots ,9$ and the number in the brackets denotes the level with respect to node ten. As a result those fields with the same level group into representations of SO(10,10). The fact that some fields are repeated indicates that they occur with the corresponding multiplicity. 
\par
The level zero fields of equation (7.5)  are the graviton, the scalar and the two form which  are just those of  the massless NS-NS sector of the IIA superstring.  The level one fields belong to the spinor representations of SO(10,10) and are the vector, the three form, five form, seven form and a nine form. The first two of these fields  are those of the massless R-R  sector of the IIA superstring while the five form and seven forms are the duals of the three and one forms respectively. As we will discuss later the nine form is associated with Romans theory. The dual graviton can be found at level two. Thus we find among these fields of the usual description of IIA supergravity their duals. 
\par
The spacetime, which is  encoded in the non-linear realisation and that  arises from this decomposition,  has the coordinates 
$$ x^a, \ \ y_a, \ \ x,\ \ x_{a_1a_2}, \ \ x_{a_1\ldots a_4}; \ \ x_{a_1\ldots a_5}, \ \ x_{a_1\ldots a_6} , \ \ x_{a_1\ldots a_6, b} , \ \ x_{a_1\ldots a_7} (2) , \ \ x_{a_1\ldots a_8} (2) , $$
$$\ x_{a_1\ldots a_8, b} (2) ,  x_{a_1\ldots a_9} (3) , \ \ x_{a_1\ldots a_{10}} (4) , \ \ x_{a_1\ldots a_7, b_1b_2}, \ \ x_{a_1\ldots a_7,b} , \ \ x_{a_1\ldots a_9,b} (4) , x_{a_1\ldots a_{10},b} (3) , 
$$
$$
\ \ x_{a_1\ldots a_9, b_1b_2} (2) , x_{a_1\ldots a_8, b_1b_2b_3} (2) , \ \ x_{a_1\ldots a_8, b_1b_2} (2) , \ \ x_{a_1\ldots a_8, (b_1b_2)}, \ \ x_{a_1\ldots a_7, b_1b_2b_3} , 
$$
$$\ \ x_{a_1\ldots a_{10}, b_1b_2} (7) , \ \ x_{a_1\ldots a_{10}, (b_1b_2)} (3) , 
 x_{a_1\ldots a_9, b_1b_2b_3} (5) , \ \ x_{a_1\ldots a_9, b_1b_2,c} (2), \ \ x_{a_1\ldots a_8,b_1\ldots b_4} (2) , 
$$
$$
\ \ x_{a_1\ldots a_8, b_1b_2b_3,c}, \ \ x_{a_1\ldots a_7, b_1\ldots b_5} , \ \ x_{a_1\ldots a_{10}, b_1b_2b_3} (7) ,  x_{a_1\ldots a_{10}, b_1b_2,c} (4) , \ \ x_{a_1\ldots a_{10}, (b_1b_2b_3)} , 
$$
$$x_{a_1\ldots a_9, b_1\ldots b_4} (4) , 
\ \ x_{a_1\ldots a_9, b_1b_2b_3, c} (3) , \ \ x_{a_1\ldots a_8, b_1\ldots b_5} (2) , \ \ x_{a_1\ldots a_7, b_1\ldots b_6}  , \ \ x_{a_1\ldots a_8, b_1\ldots b_4,c}  , \ \ldots 
\eqno(7.6)$$
where the number in brackets give the multiplicities and if there is no bracket the multiplicity is one. All the coordinates belong to irreducible representations of SL(10). The first two coordinates occur at level zero and belong to the vector representation of SO(10,10). 
\par

At level zero the  $E_{11}\otimes_s l_1$   non-linear realisation when further decomposed into representations of GL(10), as discussed above,  
 contain the massless fields in the NS-NS sector of the superstring which live on a twenty dimensional spacetime with coordinates $x^a$ and $y_a$ of equation (7.6) and the detailed equations of motion were worked out in reference [43]. The result is the  same as Siegel  theory  [44,45]. The more recent work on doubled field theory was shown to be equivalent to Siegel theory in reference [46].  The non-linear realisation up to and  including  level one  contains the above fields but also  the massless fields in the R-R sector of the superstring [47]. Indeed, it was in this paper that Siegel theory was extended to include the massless R-R fields and  so all the massless fields of IIA supergravity.  


\medskip 
{\bf 8 $E_{11}\otimes_s l_1$ non-linear realisation as a unified theory} 
\medskip 
In this section we  will explain that  the $E_{11}\otimes_s l_1$  non-linear realisation  contains all we know about maximal supergravities and so  is a unified theory. In other words  the many very different maximal supergravity theories are packaged up into this one theory. 
\par
The low level fields in the $E_{11}\otimes_s l_1$  non-linear realisation were listed above, however, it contains an infinite number of fields whose character  is largely unknown. Nonetheless in this section we will list some of the higher level fields and find out what their role in the theory is. 
In particular, it is straightforward to find all the form fields, that is, fields whose indices are totally antisymmetrised, in $D$ dimensions. These fields are  listed  in the table below  [37,49]

\medskip
\centerline {\bf Table 1. The forms
fields  in $D$ dimension with their $E_{11-D}$ representations. }
\medskip
$$\halign{\centerline{#} \cr
\vbox{\offinterlineskip
\halign{\strut \vrule \quad \hfil # \hfil\quad &\vrule Ê\quad \hfil #
\hfil\quad &\vrule \hfil # \hfil
&\vrule \hfil # \hfil Ê&\vrule \hfil # \hfil &\vrule \hfil # \hfil &
\vrule \hfil # \hfil &\vrule \hfil # \hfil &\vrule \hfil # \hfil &
\vrule \hfil # \hfil &\vrule# 
\cr
\noalign{\hrule}
D& $E_{11-D}$ & $A_a$ & $A_{a_1a_2}$ & $A_{a_1a_2a_3}$ & $A_{a_1\ldots
a_{4}}$ & $A_{a_1\ldots a_{5}}$ & $A_{a_1\ldots a_{6}}$ & $A_{a_1\ldots
a_7}$ & $A_{a_1\ldots a_8}$&\cr
\noalign{\hrule}
8&$SL(3)\otimes SL(2)$&$\bf (\bar 3,2)$&$\bf ( 3,1)$&$\bf (1,2)$&$\bf
(\bar 3,1)$&$\bf ( 3,2)$&$\bf (1,3)$&$\bf (\bar3,2)$&$\bf (3,1)$&\cr
&&&&&&&$\bf (8,1)$&$\bf (6,2)$&$\bf (15,1)$&\cr Ê&&&&&&&$$&&$
\bf (3,1)$&\cr 
&&&&&&&&&$\bf (3,3)$&\cr
\noalign{\hrule}
7&$SL(5)$&$\bf\overline{ 10}$&$\bf 5$&$\bf \bar5$&$\bf  {10}$&$\bf
24$&$\bf
\overline {40}$&$\bf 70$&-&\cr Ê&&&&&&&$\bf \overline {15}$&$\bf 45$&-&\cr
&&&&&&&&$\bf 5$&-&\cr
\noalign{\hrule}
6&$SO(5,5)$&$\bf  {16}$&$\bf 10$&$\bf \bar {16}$&$\bf 45$&$\bf 
{144}$&$\bf 320$&-&-&\cr &&&&&&&$\bf {\overline{126}}$&-&-&\cr
&&&&&&&$\bf 10$&-&-&\cr
\noalign{\hrule}
5&$E_6$&$\bf { 27}$&$\bf \overline {27}$&$\bf 78$&$\bf  {351}$&$\bf
\overline {1728}$&-&-&-&\cr Ê&&&&&&$\bf
\overline {27}$&-&-&-&\cr 
\noalign{\hrule}
4&$E_7$&$\bf 56$&$\bf 133$&$\bf 912$&$\bf 8645$&-&-&-&-&\cr
&&&&&$\bf 133$&-&-&-&-&\cr
\noalign{\hrule}
3&$E_8$&$\bf 248$&$\bf 3875$&$\bf 147250$&-&-&-&-&-&\cr
&&&$\bf 1$&$\bf 3875$&-&-&-&-&-&\cr
&&&&$\bf 248$&-&-&-&-&-&\cr\noalign{\hrule}
}}\cr}$$
\medskip

Looking at  table 1 one sees that for every form field $A_{a_1\ldots a_n}$ of rank $n$ with  $n< {D\over 2}$ indices there is  a dual field $A_{a_1\ldots a_{D-n-2}}$  that belongs to a conjugate representation of the $E_{11-D}$ algebra. One finds that the dynamics of the non-linear realisation leads to an equation that relates   the fields strengths of these  two fields through a  duality relation which is first order in derivatives [50].  See reference [16] for a review of this point. This same pattern occurred in eleven dimensions and in the ten dimensional IIA and IIB theories discussed earlier. 
\par
Thus the non-linear realisation leads to a democratic formulation in that  
 the different possible field descriptions of the degrees of freedom of the theory are present. For example,  in eleven dimensions the degrees of freedom usually encoded by the three form $A_3$ can equally well  be realised by the fields $A_6$,  It has been shown in eleven dimensions that at higher levels one finds in the  $E_{11}\otimes_s l_1$ non-linear realisation the fields $A_{3,9}, A_{3,9, 9 },A_{3,9,9,9},\dots $ or $A_{6,9}, A_{6,9, 9 },A_{6,9,9,9},\dots $ where the numbers refer to the blocks of antisymmetric indices [51]. If we were to restrict the index range to be only over nine values, as one might suppose is the case in a light-cone analysis,  then they all belong to the same representation of SO(9) and so one might think that these fields also describe the same degrees of freedom as the original $A_3$ field. Thus the $E_{11}\otimes_s l_1$ non-linear realisation provides an infinite number of ways of describing the degrees of freedom of eleven dimensional supergravity. Indeed these different possible descriptions of the particles in the theory are rotated into each other under the $E_{11}$ symmetry and so part of the $E_{11}$ symmetry can be viewed as a vast duality symmetry. Similar conclusions apply in lower dimensions. 
\par
Table 1 also contains next to top forms which are forms that have $D-1$ totally antisymmetrised SL(D) indices. We note that these fields also carry particular representations of $E_{11-D}$. Suppressing these latter indices 
such a field in $D$ dimensions  has the form $A_{a_1\ldots a_{D-1}}$;  the  corresponding field strength is of the form $F_{a_1\ldots a_{D}}$ and it should  appear in the action in the generic form 
$$
\int d^D x (\det e_\mu {}^a )F_{a_1\ldots a_{D}} F^{a_1\ldots a_{D}}
\eqno(7.7)$$
Its equation of motion is of the form $\partial_{\mu_1} ((\det e_\nu {}^a )                   F^{\mu_1\ldots \mu_{D}})=0$ and it has the solution $F_{a_1\ldots a_{D}} = m\epsilon _{a_1\ldots a_{D}}$  where $m$ is a constant. Substituting this back into the action we find a cosmological constant. Thus  the next to top forms lead to theories with cosmological constants and so the $E_{11}\otimes_s l_1$ non-linear realisation automatically contains theories with a cosmological constant which are classified by the representations of $E_{11-D}$ to which the next to top forms belong. 
\par
Supergravity theories with a cosmological constant have been studied since the discovery of the first supergravity theory. To find them one essentially  takes a  known  supergravity theory, adds by hand a cosmological constant and then tries to restore the supersymmetry by adding terms to the transformations rules and the action. It turns out that this is not possible for the eleven dimensional supergravity theory and the ten dimensional IIB theory, however, for the ten dimensional IIA theory there is a unique possibility called Romans theory [52]. For the lower dimensional maximal supergravity theories there are in fact many ways to proceed and so there are many different theories with a cosmological constant that preserve all the supersymmetries
These different theories gauge different parts of the $E_{11-D}$ symmetry and as a result such theories have become known as gauged supergravities. 
While some gauged supergravities can be obtained from ten or eleven dimensional supergravities by dimensional reduction on various manifolds, such as spheres, many have no known higher dimensional origin when viewed from the viewpoint of conventional supergravity. As a result they are not part of what is normally considered as M theory, since as we have explained,  M theory is not a theory but a set of relations between theories and for these latter theories there is no connection to the theories that are usually considered part of M theory. 
\par
There are no such next to top forms in eleven dimensions and in the IIB theory which is consistent with the fact that these theories do not have an extension to include a cosmological constant. However, the $E_{11}\otimes_s l_1$ non-linear realisation when decomposed in a way that leads to   the IIA theory in  ten  dimensions   possess a nine form, see equation (7.5),  which leads to  a deformation of the IIA theory which possess a cosmological constant [39].  This is of course Romans theory. 
\par
Examining the next to top forms in table 1 for the theories in  lower dimensions  we find that  in four dimensions they belong to   the representation of dimension 912 of $E_7$. Following our discussion just above we conclude that  these fields lead to theories with cosmological constants which are classified by the 912 representations of $E_7$.  In general the representations of the next to top forms in $D$ dimensions will   classify  all the possible gauged maximal supergravities in that dimension [37,49]. The result is in agreement with previous work carried out  over many years and based on supersymmetry [53].  Indeed this latter work used the  so called hierarchy method which  introduces some of the form fields found in  table 1.   We note that although the next to top fields do not lead to new degrees of freedom,  they clearly do lead to physical effects. 
\par
Table 1 also contains top forms, that is, forms with $D$ totally antisymmetrised indices. These will not lead to dynamical degrees of freedom but they may well lead to physical effects. We note that they will occur as the lead term in the Wess-Zumino terms in brane actions. 
\par
As we have seen the different  maximal supergravity theories arise from taking different decomposition of the $E_{11}\otimes_s l_1$ algebra and that within  a given decomposition we can also find all the gauged supergravities by taking different  next to top forms to be  is non-zero. However, there is only one $E_{11}$ algebra and only one $ l_1$ representation as such any two theories found by taking different decompositions are related to each other, that is, the fields in the different theories are related in a one to one manner and so are the coordinates [36]. It is straight forward to find the correspondence. In a given theory, or decomposition, every field component arises in the non-linear realisation from a given $E_{11}$ generator and so from a given $E_{11}$ root. To find the corresponding field component in any other theory one just has to find the one  that corresponds to the same root.  We note that the usual formulations of supergravity  contain fields that appear at the lowest levels in the non-linear realisation and one can  find that, even if the field in one theory is one of those that appears in the usual supergavity theory,   the corresponding  field in the other theory is one that is at higher level and does not  appear in the usual description of this other  supergravity theory. As similar argument applies to the coordinates in the different theories. 
\par
The correspondence between the different theories is especially  interesting to examine for the gauged supergravity and in particular how  a non-zero  next to top form,  which is therefore responsible for the cosmological constant,  is mapped into another field in a different theory. Indeed one can find what field in the eleven dimensional theory corresponds to a given  next to top form in lower dimensions and as a result find the eleven dimensional origin of all gauged supergravities.  We now give two examples of this procedure. Let us first consider Romans theory. As we discussed above this theory arises from the nine form of equation (7.5). However, rather than tracing to what field   the  $E_{11}$ root of this field corresponds to in eleven dimensions it is simpler, in this case,  to  carry out the dimensional reduction from eleven dimensions directly. The $E_{11}\otimes_s l_1$  non-linear realisation when decomposed into representations of GL(11)   leads to the eleven dimensional fields [23, 55]
$$
h_{\hat a}{}^{\hat b}\ (0), A_{\hat a_1\hat a_2\hat a_3}\ (1),  
A_{\hat a_1\ldots \hat a_6}\ (2), h_{\hat a_1\ldots\hat a_8,\hat b}\ (3),
$$
$$ 
A_{\hat a_1\ldots\hat a_9,\hat b_1\hat b_2\hat b_3}\ (4), 
A_{\hat a_1\ldots\hat a_{11},\hat b}\ (4), A_{\hat a_1\ldots\hat
a_{10},(\hat b_1\hat b_2 )}\ (4), 
\ldots 
\eqno(7.8)$$
where $\hat a, \hat b, \ldots =1,\ldots 11$ and the number in the brackets is the level. 
Carrying  out the dimensional reduction to ten dimensions by hand we get the IIA fields and it is easy to see that the nine form in ten dimensions arises from the level four field $A_{a_1\ldots a_{10} , bc}$ which is antisymmetric in its $a_1, \ldots ,a_{10}$ indices but symmetric in the $b, c$ indices. Indeed the nine form arises  as $A_{a_1\ldots a_{9}11 , 11 11}$. We note that this level four field is in a part of the $E_{11}\otimes_s l_1$ non-linear realisation  that is beyond the eleven dimensional supergravity theory.  Thus we have found the eleven dimensional origin of Romans theory,  something that could not have been found using conventional supergravity techniques.   
\par
Our second example concerns four dimensions where the next to top fields have the form $A_{a_1a_2a_3\bullet} $ where $\bullet$ refers to the 912- dimensional representation of $E_7$ to which this field belongs. Decomposing this representation to representations of SL(8) we find that [37] 
$$
912 \to 420 \oplus \overline {420} \oplus 36 \oplus \overline {36} 
\eqno(7.9)$$
which correspond to the tensors 
$$
 \phi^{I_1\ldots I_3}{}_{ J}\oplus \phi_{I_1\ldots I_3}{}^{ J}\oplus \phi_{(I_1 I_2)}
\oplus \phi^{(I_1 I_2)}
\eqno(7.10)$$
where $I,J,\ldots =1,2,\ldots 8$. The reader can find a detailed account of how the next to top fields arise from the eleven dimensional fields by dimensional reduction in reference [37]. 
\par
In particular, let us look for a  theory that has a  SO(8) gauging of the $E_7$ symmetry. This is achieved if we take the next to top field $A_{a_1a_2a_3\bullet} $ to be a singlet under SO(8). Looking at the above representations of SL(8) of equation (7.9), and decomposing them into SO(8) representations,  we see that there are only two singlets, one in the $36$ and the other in the $\bar {36}$. 
The  fields of the 36-dimensional representation  arises from the eleven dimensional fields $A_3$ and $A_{9,6}$ 
which is consistent with the known eleven dimensional origin of  this four dimensional gauged supergravity that arises from dimensional reduction on a seven sphere [37]. The  SO(8) singlet in the other 36-dimensional representation arises from the eleven dimensional fields $A_{10,1,1}$ 
and $A_{10,7,7}$ which shows that this  four dimensional gauged supergravity has no  eleven-dimensional supergravity origin, but of course it does have an eleven dimensional origin in the $E_{11}\otimes_s l_1$ non-linear realisation [37]. 
\par
While it is clear from the above discussion that the $E_{11}\otimes_s l_1$  non-linear realisation does contain all the gauged supergravities,  it would be interesting  to show in detail how the dynamical equations that encode   their origin follow from the non-linear realisation. 
\par
The spacetime contained in the non-linear realisation also contains an infinite number of coordinates whose detailed form is only known at low levels. However, just as for the fields one can find all the form coordinates, that is, coordinates that have one block of antisymmetrised indices. The result for the corresponding generators in the $l_1$ representation are  given in the following table. [27, 54, 26].  
\medskip 
{\centerline{\bf {Table 2. The form generators  in the $l_1$ representation  in D
dimensions}}}
\medskip
$$\halign{\centerline{#} \cr
\vbox{\offinterlineskip
\halign{\strut \vrule \quad \hfil # \hfil\quad &\vrule Ê\quad \hfil #
\hfil\quad &\vrule \hfil # \hfil
&\vrule \hfil # \hfil Ê&\vrule \hfil # \hfil &\vrule \hfil # \hfil &
\vrule \hfil # \hfil &\vrule \hfil # \hfil &\vrule \hfil # \hfil &
\vrule \hfil # \hfil &\vrule#
\cr
\noalign{\hrule}
D&G&$Z$&$Z^{a}$&$Z^{a_1a_2}$&$Z^{a_1\ldots a_{3}}$&$Z^{a_1\ldots a_
{4}}$&$Z^{a_1\ldots a_{5}}$&$Z^{a_1\ldots a_6}$&$Z^{a_1\ldots a_7}$&\cr
\noalign{\hrule}
8&$SL(3)\otimes SL(2)$&$\bf (3,2)$&$\bf (\bar 3,1)$&$\bf (1,2)$&$\bf
(3,1)$&$\bf (\bar 3,2)$&$\bf (1,3)$&$\bf (3,2)$&$\bf (6,1)$&\cr
&&&&&&&$\bf (8,1)$&$\bf (6,2)$&$\bf (18,1)$&\cr Ê&&&&&&&$\bf (1,1)$&&$
\bf
(3,1)$&\cr Ê&&&&&&&&&$\bf (6,1)$&\cr
&&&&&&&&&$\bf (3,3)$&\cr
\noalign{\hrule}
7&$SL(5)$&$\bf 10$&$\bf\bar 5$&$\bf 5$&$\bf \overline {10}$&$\bf 24$&$\bf
40$&$\bf 70$&-&\cr Ê&&&&&&$\bf 1$&$\bf 15$&$\bf 50$&-&\cr
&&&&&&&$\bf 10$&$\bf 45$&-&\cr
&&&&&&&&$\bf 5$&-&\cr
\noalign{\hrule}
6&$SO(5,5)$&$\bf \overline {16}$&$\bf 10$&$\bf 16$&$\bf 45$&$\bf \overline
{144}$&$\bf 320$&-&-&\cr &&&&&$\bf 1$&$\bf 16$&$\bf 126$&-&-&\cr
&&&&&&&$\bf 120$&-&-&\cr
\noalign{\hrule}
5&$E_6$&$\bf\overline { 27}$&$\bf 27$&$\bf 78$&$\bf \overline {351}$&$\bf
1728$&-&-&-&\cr Ê&&&&$\bf 1$&$\bf \overline {27}$&$\bf 351$&-&-&-&\cr
&&&&&&$\bf 27$&-&-&-&\cr
\noalign{\hrule}
4&$E_7$&$\bf 56$&$\bf 133$&$\bf 912$&$\bf 8645$&-&-&-&-&\cr
&&&$\bf 1$&$\bf 56$&$\bf 1539$&-&-&-&-&\cr
&&&&&$\bf 133$&-&-&-&-&\cr
&&&&&$\bf 1$&-&-&-&-&\cr
\noalign{\hrule}
3&$E_8$&$\bf 248$&$\bf 3875$&$\bf 147250$&-&-&-&-&-&\cr
&&$\bf1$&$\bf248$&$\bf 30380$&-&-&-&-&-&\cr
&&&$\bf 1$&$\bf 3875$&-&-&-&-&-&\cr
&&&&$\bf 248$&-&-&-&-&-&\cr
&&&&$\bf 1$&-&-&-&-&-&\cr
\noalign{\hrule}
}}\cr}$$
\medskip
From the above table for the generators in the $l_1$ representation we can read off the coordinates in the spacetime that occurs in the non-linear realisation. At level zero we find the coordinates of the spacetime in $D$ dimensions that we are familiar with. However,  at level one we find   coordinates which are scalars under the SL(D) transformations of our usual spacetime, and so also Lorentz transformations,  but belong to non-trivial representations of $E_{11-D}$. In particular, they belong to the 
 $$
10,\quad\overline {16}, \quad \overline {27},\quad  56, \quad {\rm  and}\quad 
248\oplus 1 , \quad {\rm of }\quad SL(5),\quad  SO(5,5),\quad 
E_6, \quad  E_7\quad {\rm  and}\  E_8 
\eqno(1.2)$$
 for $  D= 7,6,5,4$ and $3$    dimensions  respectively [27,54].
\par
The coordinates play an essential role in the derivation of the equations of motion and one can not find invariant equations without them. We see from equation (6.18) that if the first ($l_1$) index of  the Cartan form is of level one, that is, $G^{a_1a_2}{}_{, \bullet}$ then it varies into a Cartan form with a first index that is a usual spacetime index and so it  contains derivatives with respect to the usual coordinates. Hence, the terms in the equations of motion with derivatives with respect to the usual derivatives and the higher level derivatives mix. As we explained above, to find the gauged supergravities in the non-linear realisation one had to take some next to top forms to be non-zero  but one also has to take the fields to depend on the higher level coordinates in a non-trivial way [38]. Nonetheless , the physical role that the higher level coordinates play is not  at all well understood. However, the very fact that the final truncated equations of motion are precisely those of the maximal supergravity theories and that the higher level coordinates are essential to find this result suggests that they play an important role in a way we have yet to understand. Given the unfamiliar nature of the higher level coordinates rather than give in to the  temptation to invent mathematical tricks to try to eliminate them it may be better to try to find their underlying physical meaning.

\medskip
{\bf 8 Discussion}
\medskip
We have reviewed  the theory of non-linear realisations and explained how it leads to dynamical equations of motion. We also have recalled  how non-linear realisations played a key role in the introduction of symmetry and spontaneous symmetry breaking into particle physics. The theory of Kac-Moody algebras was   briefly discussed as well as  the construction of the Kac-Moody algebra $E_{11}$ together with its vector representation $l_1$. The non-linear realisation of $E_{11}\otimes_s l_1$ was constructed and the dynamics that it implies was derived. It leads to an $E_{11}$ invariant field theory that has an infinite number of fields which depend on a spacetime that has an infinite number of coordinates. However, the uniquely determined dynamics  agrees precisely with the equations of motion of the eleven dimensional supergravity theory when we restrict the fields to be those at lowest levels and the coordinates to be just those of our usual spacetime. 
\par
The dynamics in eleven dimensions was derived by taking a decomposition of $E_{11}$ into its GL(11) subgroup. However, by taking decompositions of    $E_{11}$ into different subalgebras we found all the maximal supergravity theories in ten and less dimensions. Although the detailed calculations have only been carried out in five dimensions,  it is inevitable that the dynamics of the non-linear realisation of $E_{11}\otimes_s l_1$  in the different decompositions will agree with the equations of motion of the corresponding supergravity theories, in the same sense as just mentioned above. We also explained how the maximal gauged supergravities are automatically included in the $E_{11}\otimes_s l_1$ non-linear realisation. As result the $E_{11}\otimes_s l_1$ non-linear realisation is a unified theory in the sense that it contains all the maximal supergravities. It also follows from the way the different theories arise from the  different decompositions that all   the  theories  derived from the non-linear realisation are completely equivalent in that  the coordinates and fields are just rearranged from one theory to another according  to the different   decompositions of $E_{11}\otimes_s l_1$ being used. We note that eleven dimensions does not play the preferred role as it does in M theory as all the theories are on an equal footing. 
\par
The maximal type II supergravity theories were thought to be the complete low energy effective actions for the type II superstrings. However,   as the $E_{11}\otimes_s l_1$ 
non-linear realisation  contains all these theories in one unified structure it is difficult not to believe that the conjecture of [23,24],  namely  that the $E_{11}\otimes_s l_1$ 
non-linear realisation is the low energy effective action for the type II superstrings. This theory contains many effects which are beyond those found in the supergavity theories and it will be very interesting to find out in detail what these effects are.  
Indeed this work provides  a starting point from which to more systematically consider what is the underlying theory of strings and branes. 
\par
One obvious question is whether the higher level fields lead to additional degrees of freedom beyond those found at low levels which are just those of the maximal supergravity theories. While the answer to this question is not known for sure it is likely that this is not the case. As one examines the fields at levels just above the supergravity fields  one does not find fields that lead to new degrees of freedom. Also, in eleven dimensions,  all fields that do not have blocks of ten of eleven antisymmetrised indices have been classified [51]. One finds an infinite number of such fields, however,  they are the fields of  the supergravity theories plus  fields  that are dual to these fields. One expects that these dual fields satisfy first order duality relations and so not lead to any new degrees of freedom. Thus if one adopted a light-cone description which takes into account only objects which carry indices ranging  over  nine different values then one might, perhaps naively, expect that only the above fields would lead to dynamical degrees of freedom. As a result one expects to find only the usual degrees of freedom of the eleven dimensional supergravity theories. A possible exception is the dependence of the fields on the additional coordinates which could lead to new degrees of freedom as it does in higher spin theories [56]. The same conclusion applies to all the maximal super gravity theories. 
\par
As we have explained a non-linear realisation provides a very direct path from the  algebra used in the non-linear realisation to the dynamics. 
In the case of the $E_{11}\otimes_s l_1$  non-linear realisation it provides a direct path from the $E_{11}$ Dynkin diagram to the equations of motion of the maximal supergravities. We note that the dynamics of this  non-linear realisation is uniquely determined, at least at low levels. 
The only assumptions we make are  that  we use the vector representation of $E_{11}$ to build the semi-direct product algebra and that we require the smallest number space time derivatives which leads to non-trivial dynamics. It is amusing to note that one can uniquely derive Einstein's theory of general relativity in this way, that is, it is contained in this sense in the Dynkin diagram of $E_{11}$. 
\par
In this review we have focused entirely on the bosonic sector of the supergravity fields. One can introduce fermions as fields that transform under $I_c(E_{11}))$ [57] following  a similar procedure [58] to that carried out  in the context of the $E_{10}$ approach. 
\par
The symmetries of the non-linear realisation do not include the local symmetries of gauge and general coordinate transformations. However, the equations of motion that follow from the non-linear realisation are unique and they turn out to be general coordinate and gauge invariant. It would be interesting to see if this phenomenon persists at higher levels and why it is that  these local   symmetries arise in this way. 
\par
Although the coordinates beyond those of the usual spacetime must be truncated out of the equations of the $E_{11}\otimes_s l_1$ non-linear realisation to find the equations of motion we are used to, they play an essential role in  the way the equations of motion were derived. Indeed they are crucial for  the $E_{11}$ symmetry and one could not derive these equations without them.  We should think of these extra coordinates as leading to physical effects, indeed they are required for the gauged supergravities. It is very unlikely that our usual notion of spacetime 
survives in a fundamental theory of physics and in particular in the underlying theory of strings and branes. One can think of  the infinite dimensional spacetime that appears in the  $E_{11}\otimes_s l_1$ non-linear realisation  as  a kind of low energy effect theory of spacetime that represents the properties of spacetime before it is replaced by more fundamental  degrees of freedom. This can be thought of as  analogous  to the   low energy effective actions which do not contain  the fields that correspond to all the degrees of freedom in the underlying theory but  
only the fields corresponding to degrees of freedom which have a low mass compared to the scale being considered. 
 The problem of how to eliminate all the higher level coordinates in the applications we are used to is a problem whose resolution demands a physical as well as a mathematical idea. Truncating the coordinates breaks the $E_{11}$ symmetry,  however when one better understands the role of the extra coordinates this breaking may appear as some kind of spontaneous rather than explicit symmetry breaking. As we recalled from the history of particle physics,  one can not hope to solve all the problems in one go.

\medskip
{\bf {Acknowledgements}}
\medskip
I wish to thank Nikolay Gromov and Alexander Tumanov for very helpful discussions   I also wish to thank the SFTC for support from Consolidated grant number ST/J002798/1.

\medskip
{\bf {References}}
\medskip
\item{[0]} S.Weinberg, {\it The quantum theory of fields}, vol 1, Cambridge University Press, 1995; M. Veltman, {\it Facts and Mysteries in Elementary Particle Physics}, World Scientific 2003; Contributions in {\it Shelter Island II} edited by R. Jackiw, N. Khuri, S. Weinberg and E. Witten, MIT Press, 1985; C. N. Yang, {\it Selected Papers 1945-1980}, W. H. Freeman and Company, 1983; Contributions of S. Coleman, M. Gell-mann, ÊS. Glashow Êand B. Zumino in {\it Hadrons and their Interactions}, 1967 International School of Physics Ettore Majorana, edited by A Zichichi, Academic Press 1968. The reader may like to read the wonderful lectures of Zumino that explain how the results  of current algebra can be found in a very simple way from the non-linear realisation of $SU(2)\otimes SU(2)$. 
\item{[1]} V. Alessandrini, {\it A General approach to dual multiloop diagrams}, Nuov. Cim. {\bf 2A} (1971) 321; V. Allessandrini and D. Amati, {\it Properties of dual multiloop amplitudes}, Nuov. Cim. {\bf 4A} (1971) 793; C. Lovelace, {\it M-loop generalized veneziano formula.},
Phys. Lett. {\bf 32} (1970) 703. 
\item{[2]} V. Alessandrini, D. Amati, M. LeBellac and D.I. Olive, {\it The operator approach to dual multiparticle theory},
Physics Reports {\bf 1C} (1971) 170; as well as the additional
supplement
by D. Olive which appears immediately after this article.
\item{[3]}A. Neveu and J. Scherk, {\it Gauge invariance and uniqueness of the renormalisation of dual models with unit intercept}, Nucl. Phys.
{\bf {B36}} (1972) 155.
\item {[4]} T. Yoneya, {\it Connection of Dual Models to Electrodynamics and Gravidynamics}, Progr. Theor. Phys. {\bf {51}}, (1974)
1907; ÊJ. Scherk and J. Schwarz, {\it Dual Models for Nonhadrons}, Nucl. Phys.{\bf {B81}} Ê(1974) 118.
\item{[5]} C. Campbell and P. West,
{\it $N=2$ $D=10$ nonchiral
supergravity and its spontaneous compactification.}
Nucl.\ Phys.\ {\bf B243} (1984) 112;  M. Huq and M. Namazie,
{\it Kaluza--Klein supergravity in ten dimensions},
Class.\ Q.\ Grav.\ {\bf 2} (1985) 597;  F. Giani and M. Pernici,
{\it $N=2$ supergravity in ten dimensions},
Phys.\ Rev.\ {\bf D30} (1984) 325.
\item{[{6}]} J, Schwarz and P. West,
{\it ÊSymmetries and Transformation of Chiral
$N=2$ $D=10$ Supergravity},
Phys. Lett. {\bf 126B} (1983) 301.
\item {[{7}]} P. Howe and P. West,
{\it The Complete $N=2$ $D=10$ Supergravity},
Nucl.\ Phys.\ {\bf B238} (1984) 181.
\item {[{8}]} J. Schwarz,
{\it Covariant Field Equations of Chiral $N=2$
$D=10$ Supergravity},
Nucl.\ Phys.\ {\bf B226} (1983) 269.
\item{[9]} E. Cremmer, B. Julia and J. Scherk, {\it Supergravity Theory in Eleven-Dimensions}, Phys.
Lett. {\bf 76B} (1978) 409.
\item{[10]} E. Cremmer and B. Julia,
{\it The $N=8$ supergravity theory. I. The Lagrangian},
Phys.\ Lett.\ {\bf 80B} (1978) 48.
\item {[11]} B.\ Julia, {\it ÊGroup Disintegrations},
in {\it Superspace
Supergravity}, p.\ 331, Êeds.\ S.W.\ Hawking Êand M.\ Ro\v{c}ek,
Cambridge University Press (1981);
ÊE. Cremmer, {\it Dimensional Reduction In
Field Theory And Hidden Symmetries In Extended Supergravity}, ÊPublished
in Trieste Supergravity School 1981, 313; Ê{\it Supergravities In 5
Dimensions}, in {\it Superspace
Supergravity}, p.\ 331, Êeds.\ S.W.\ Hawking Êand M.\ Ro\v{c}ek,
Cambridge University Press (1981).
\item{[12]}A. Font, L. Ibanez, D. Lust and F. D. Quevedo, {\it Strong - weak coupling duality and nonperturbative effects in string theory}, Phys.
Lett. Ê{\bf 249B} (1990) 35;  S.J. Rey, {\it The Confining Phase Of Superstrings And Axionic Strings}, Phys. Rev. {\bf D43} (1991) 526.
\item{[13]} C.M. Hull and P.K. Townsend,
{\it Unity of superstring Êdualities},
Nucl.\ Phys.\ {\bf B438} (1995) 109, hep-th/9410167.
\item{[14]} P. Townsend, {\it The eleven dimensional
supermembrane revisited}, Phys.\ Lett.\ {\bf 350B} (1995) 184,
arXiv:hep-th/9501068;  E. Witten, {\it String theory dynamics in various
dimensions},
Nucl.\ Phys.\ {\bf B443} (1995) 85, ÊarXiv:hep-th/9503124.
\item{[15]} S. Coleman, J. Wess and ÊB. Zumino,
{\it ÊStructure of Phenomenological Lagrangians. 1},
Phys.Rev. {\bf 177} (1969) 2239; C. Callan, S.
Coleman, J. Wess and B. Zumino, 
{\it Structure of phenomenological Lagrangians. 2},  Phys. Rev. {\bf 177}
(1969)  2247.
\item{[16]} P. West, {\it Introduction to Strings and Branes}, Cambridge University Press, 2012. 
\item{[17]}A. Salam and J. Strathdee, {\it Nonlinear realizations. 1: The Role of Goldstone bosons}, Phys. Rev. {\bf 184} (1969)
 1750, C. Isham, A. Salam and J. Strathdee, {\it Spontaneous,
breakdown of conformal symmetry}, Phys. Lett. {\bf 31B} (1970) 300.
\item{[18]} A. Borisov and V. Ogievetsky, {\it Theory of
dynamical affine and conformal symmetries as gravity theory Êof
ÊÊthe gravitational field}, ÊTheor.\ Math.\ Phys.\ Ê{\bf 21} (1975)
1179; ÊV. Ogievetsky, Ê{\it ``Infinite-dimensional algebra of
general Êcovariance group as the closure of the finite dimensional
algebras Êof conformal and linear groups"},
Nuovo. Cimento, {\bf 8} (1973) 988.
\item{[19]} D.V. Volkov, {\it Phenomological Lagrangians},  Sov. J. Part. Nucl. {\bf 4} (1973) 3. 
\item{[20]} V. Kac, {\it Graded Lie algebras and Symmetric Spaces},
Funct. Anal. Appl., Ê{\bf 2} (1968) 183; V. Kac, {\it Infinite Dimensional
Lie Algebras}, Birkhauser, 1983.
\item {[21]} R. Moody, {\it A New Class of Lie Algebras}, J.
Algebra {\bf 10} (1968) 211.
\item{[22]} P. West, {\it Hidden Superconformal Symmetry in M Theory}, JHEP 0008 (2000) 007, hep-th/0005270.
\item{[23]} P. West, {\it $E_{11}$ and M Theory}, Class. Quant.  
Grav.  {\bf 18}, (2001) 4443, hep-th/ 0104081. 
\item{[24]} P. West, {\it $E_{11}$, SL(32) and Central Charges},
Phys. Lett. {\bf B 575} (2003) 333-342,  hep-th/0307098. 
\item{[25]}  A. Kleinschmidt and P. West, {\it  Representations of G+++
and the role of space-time},  JHEP 0402 (2004) 033,  hep-th/0312247.
\item{[26]} P. Cook and P. West, {\it Charge multiplets and masses
for E(11)}, ÊJHEP {\bf 11} (2008) 091, arXiv:0805.4451.
\item{[27]} P. West,  {\it $E_{11}$ origin of Brane charges and U-duality
multiplets}, JHEP 0408 (2004) 052, hep-th/0406150. 
\item{[28]} T.   Nutma,    SimpLie,    a   simple   program   for   Lie   algebras, https://code.google.com/p/simplie/.
\item{[29]} F. Englert and L. Houart, {\it G+++ Invariant Formulation of Gravity and M-Theories: Exact BPS Solutions }, JHEP0401:002,2004,  arXiv:hep-th/0311255. 
\item{[30]} Abdus Salam, {\it Gauge unification of fundamental forces}, Nobel lecture on December 8, 1979, Reviews of Modern Physics, Vol 52, no 3,  (1980), 306.
\item{[31]}  P. West, {\it Generalised Geometry, eleven dimensions
and $E_{11}$}, JHEP 1202 (2012) 018, arXiv:1111.1642.  
\item{[32]} A. Tumanov and P. West, {\it Generalised vielbeins and non-linear realisations }, JHEP 1410 (2014) 009,  arXiv:1405.7894. 
\item{[33]} A. Tumanov and P. West, {\it E11 must be a symmetry of strings and branes },  Phys. Lett. {\bf  B759 }Ê(2016),  663, arXiv:1512.01644. 
\item{[34]} A. Tumanov and P. West, {\it E11 in 11D}, Phys.Lett. B758 (2016) 278, arXiv:1601.03974. 
\item{[35]} I. Schnakenburg and  P. West, {\it Kac-Moody   
symmetries of
IIB supergravity}, Phys. Lett. {\bf B517} (2001) 421, hep-th/0107181.
\item{[36]} P. West, {\it The IIA, IIB and eleven dimensional theories 
and their common
$E_{11}$ origin}, Nucl. Phys. B693 (2004) 76-102, hep-th/0402140. 
\item{[37]}  F. ÊRiccioni and P. West, {\it
The $E_{11}$ origin of all maximal supergravities}, ÊJHEP {\bf 0707}
(2007) 063; ÊarXiv:0705.0752.
\item{[38]} ÊF. Riccioni and P. West, {\it E(11)-extended spacetime
and gauged supergravities},
JHEP {\bf 0802} (2008) 039, ÊarXiv:0712.1795.
\item{[39]} A.~Kleinschmidt, I.~Schnakenburg and P.~West, {\it Very-extended Kac-Moody algebras and their interpretation at low  levels}, Class.\ Quant.\ Grav.\  {\bf 21} (2004) 2493 [arXiv:hep-th/0309198].; P.~West, {\it E(11), ten forms and supergravity},  JHEP {\bf 0603} (2006) 072,  [arXiv:hep-th/0511153]. 
\item{[40]}P.~West, {\it E(11), ten forms and supergravity},  JHEP {\bf 0603} (2006) 072,  [arXiv:hep-th/0511153]. 
\item{[41]} P. Messen and T. Ortin, {\it An SL(2,Z) multiplet of nine
-dimensional type II supergravity theories}, Nucl. Phys. B {\bf 541}
(1999) 195, hep-th/9806120; 
G. Dall'Agata, K. Lechner and M. Tonin, {\it D=10, N=IIB supergravity;
Lorentz-invariant actions and duality}, JHEP {\bf 9807} (1998) 017,
hep-th/9806140;
E. Bergshoeff, U. Gran and D. Roest, {\it Type IIB seven-brane solutions
from nine-dimensional domain walls},  
  Class. Quant. Grav. {\bf 19} (2002) 4207, hep-th/0203202.
\item{[42]} E.~A.~Bergshoeff, M.~de Roo, S.~F.~Kerstan and F.~Riccioni,
  {\it IIB supergravity revisited},JHEP {\bf 0508} (2005) 098
  [arXiv:hep-th/0506013]; E.~A.~Bergshoeff, M.~de Roo, S.~F.~Kerstan, T.~Ortin and F.~Riccioni, {\it IIA ten-forms and the gauge algebras of maximal supergravity theories},  JHEP {\bf 0607} (2006) 018
  [arXiv:hep-th/0602280]. 
\item{[43]} P. West, {\it E11, generalised space-time and IIA string
theory}, 
 Phys.Lett.B696 (2011) 403-409,   arXiv:1009.2624.
\item{[44]} W. Siegel, {\it Two vielbein formalism for string inspired axionic gravity},   Phys.Rev. D47 (1993) 5453,  hep-th/9302036; 
\item{[45]} W. Siegel,{\it Superspace duality in low-energy superstrings}, Phys.Rev. D48 (1993) 2826-2837, hep-th/9305073; 
{\it Manifest duality in low-energy superstrings},  
In *Berkeley 1993, Proceedings, Strings '93* 353,  hep-th/9308133. 
\item{[46]} O. Hohm and S.  Kwak, {\it Frame-like Geometry of Double Field Theory},   J.Phys.A44 (2011) 085404, arXiv:1011.4101. 
\item{[47]} A. Rocen and P. West,  {\it E11, generalised space-time and
IIA string theory;  the R-R sector}, in {\it Strings, Gauge fields and the Geometry 
behind:The Legacy of Maximilian Kreuzer} edited by  Anton Rebhan, Ludmil Katzarkov,  Johanna Knapp, Radoslav Rashkov, Emanuel Scheid, World Scientific, 2013, arXiv:1012.2744.
\item{[48]} P. West, {\it  E11, Generalised space-time and equations of motion in four dimensions}, JHEP 1212 (2012) 068, arXiv:1206.7045. 
\item{[49]} E. Bergshoeff, I. De Baetselier and  T. Nutma, {\it 
E(11) and the Embedding Tensor},  JHEP 0709 (2007) 047, arXiv:0705.1304. 
\item{[50]} F. ÊRiccioni, ÊD. ÊSteele and P. West, {\it The E(11)
origin of all maximal supergravities - the hierarchy of field-strengths}
ÊÊJHEP {\bf 0909} (2009) 095, arXiv:0906.1177.
\item{[51]} F. Riccioni and P. West, {\it Dual fields and $E_{11}$},   Phys.Lett.B645 (2007) 286-292,  hep-th/0612001; F. Riccioni, D. Steele and P. West, {\it Duality Symmetries and $G^{+++}$ Theories},  Class.Quant.Grav.25:045012,2008,  arXiv:0706.3659. 
\item{[52]}  L. J. Romans, {\sl Massive $N=2A$ supergravity in ten
  dimensions}, Phys. Lett. {\bf B 169} (1986) 374.
\item{[53]}  See for example, B.~de Wit, H.~Samtleben and M.~Trigiante,
  {\it The maximal D = 5 supergravities},
  Nucl.\ Phys.\  B {\bf 716} (2005) 215, arXiv:hep-th/0412173; 
 B.~de Wit and H.~Samtleben,
  {\it Gauged maximal supergravities and hierarchies of nonabelian vector-tensor systems}, Fortsch.\ Phys.\  {\bf 53} (2005) 442, arXiv:hep-th/0501243 and references therein.
\item{[54]} P. West, {\it Brane dynamics, central charges and
$E_{11}$}, JHEP 0503 (2005) 077, hep-th/0412336. 
\item{[55]} P. West, {\it Very Extended $E_8$ and $A_8$ at low
levels, Gravity and Supergravity}, Class.Quant.Grav. {\bf 20} (2003)
2393, hep-th/0212291.
\item{[56]} P. West, {\it $E_{11}$ and Higher Spin Theories}, 
Phys.Lett.B650:197-202,2007,  hep-th/0701026.  
\item{[57]}D. Steele and P. West, {\it E11 and Supersymmetry}, JHEP 1102 (2011) 101,  arXiv:1011.5820.
\item{[58]} S. de Buyl, M. Henneaux and  L. Paulot, {\sl Extended E8
Invariance of 11-Dimensional Supergravity} JHEP {\bf 0602} (2006) 056
{\tt hep-th/05122992};  T. Damour, A. Kleinschmidt qand  H. Nicolai {\sl
Hidden symmetries and the fermionic sector of eleven-dimensional
supergravity} Phys. Lett. B {\bf 634} (2006) 319 {\tt hep-th/0512163}; S. de Buyl, M. Henneaux and  L. Paulot {\sl Hidden
Symmetries and Dirac Fermions}Ê Class. Quant. Grav. {\bf 22} (2005) 3595
{\tt hep-th/0506009}.


\end

\item{[17]} A. Tumanov and P. West, {\it E11 and exceptional field theory }, 
 arXiv:1507.08912.  

\item{[20]} P. West, {\it Dual gravity and E11  },  arXiv:1411.0920. 
\item{[21]} P. West, {\it Generalised Space-time and Gauge Transformations},
JHEP 1408 (2014) 050,  arXiv:1403.6395.  

\item{[26]}  O. Hohm and  H. Samtleben, {\it Exceptional Form of D=11 Supergravity}, Phys. Rev. Lett. 111 (2013)  231601, arXiv:1308.1673; 
 {\it Exceptional Field Theory I: $E_{6(6)}$ covariant Form of M-Theory and Type IIB},  Phys. Rev. D 89, (2014) 066016 , arXiv:1312.0614;  {\it Exceptional Field Theory II: E$_{7(7)}$} ,  arXiv:1312.0614; {\it   Exceptional Field Theory III: E$_{8(8)}$},  Phys. Rev. D 90, (2014) 066002, arXiv:1406.3348.

\end